\documentclass[a4paper]{article}  

\def\comment#1{} 
\def\umlaut{\"} 

\def\journalfont{\rm}         
\def\jou#1{{\journalfont #1\ }}
\def\joudef#1#2{\def #1{\jou{\ignorespaces #2}}}

\joudef{\aaa}    { Astron.\ Astrophys.}
\joudef{\aip}    { Adv.\ Phys.}
\joudef{\adm}    { adv.\ math.}
\joudef{\aihpa}  { Ann.\ Inst.\ H.\ Poincar\'e A}
\joudef{\am}     { Ann.\ Math.}
\joudef{\apny}   { Ann.\ Phys.\ (N.Y.)}
\joudef{\apj}    { Astrophys.\ J.}
\joudef{\cjp}    { Can.\ J.\ Phys.}
\joudef{\cmp}    { Commun.\ Math.\ Phys.}
\joudef{\cqg}    { Class.\ Quantum Grav.}
\joudef{\grg}    { Gen.\ Rel.\ Grav.}
\joudef{\gp}      {Geom.\ and Phys.}
\joudef{\ijmpd}  { Int.\ J.\ Mod.\ Phys.\ D}
\joudef{\ijtp}   { Int.\ J.\ Theor.\ Phys.}
\joudef{\invm}   { Invent.\ Math.}
\joudef{\jm}     { J.\ Math.}
\joudef{\jmaa}   { J.\ Math.\ Anal.\ Appl.}
\joudef{\jmp}    { J.\ Math.\ Phys.}
\joudef{\jpa}    { J.\ Phys.\ A}
\joudef{\mnras}  { Mon.\ Not.\ R.\ Ast.\ Soc.}
\joudef{\mpla}   { Mod.\ Phys.\ Lett.\ A} 
\joudef{\nature} { Nature}
\joudef{\nc}     { Nuovo Cim.}
\joudef{\ncb}    { Nuovo Cim. B}
\joudef{\npb}    { Nuc.\ Phys.\ B}
\joudef{\ph}     { Physica}
\joudef{\pla}    { Phys.\ Lett. A}
\joudef{\plb}    { Phys.\ Lett. B}
\joudef{\pr}     { Phys.\ Rev.}
\joudef{\prd}    { Phys.\ Rev.\ D}
\joudef{\prep}   { Phys.\ Rep.}
\joudef{\prl}    { Phys.\ Rev.\ Lett.}
\joudef{\prsla}  { Proc.\ Roy.\ Soc.\ Lond.\ A}
\joudef{\ptp}    { Prog.\ Theor.\ Phys.}
\joudef{\ptps}   { Prog.\ Theor.\ Phys.\ Suppl.}
\joudef\rmp      { Rev.\ Mod.\ Phys.}
\joudef\spj      { Sov.\ Phys.\ JETP}

\catcode`@=11

\def\eqalign#1{\null\,\vcenter{\openup\jot\m@th
  \ialign{\strut\hfil$\displaystyle{##}$&$\displaystyle{{}##}$\hfil
      \crcr#1\crcr}}\,}
\def\meqalign#1{\null\,\vcenter{\openup\jot\m@th
  \ialign{\strut\hfil$\displaystyle{##}$&&$\displaystyle{{}##}$\hfil
      \crcr#1\crcr}}\,}
\catcode`@=12   

\newdimen\arrayruleHwidth
\makeatletter
\def\Hline{\noalign{\ifnum0=`}\fi\hrule \@height \arrayruleHwidth
  \futurelet \@tempa\@xhline}
\makeatother

\newcommand\thickbaselines{\baselineskip=20pt\lineskip=3pt\lineskiplimit=3pt}
\catcode`@=11
\def\cases#1{\left\{\,\vcenter{\thicknormalbaselines\m@th
             \ialign{$##\hfil$&\quad##\hfil\crcr#1\crcr}}\right.}
\def\matrix#1{\null\,\vcenter{\thickbaselines\m@th
    \ialign{\hfil$##$\hfil&&\quad\hfil$##$\hfil\crcr
      \mathstrut\crcr\noalign{\kern-\baselineskip}
      #1\crcr\mathstrut\crcr\noalign{\kern-\baselineskip}}}\,} 
\catcode`@=12   

\newcommand{\eprint}{\textsf} 

\newcommand\be{\begin{equation}} \newcommand\ee{\end{equation}} 
\newcommand\bd{\begin{displaymath}}\newcommand\ed{\end{displaymath}}

\newcommand\Tr{\mathop{\rm Tr}\nolimits}
\newcommand\ts\textstyle

\def\undersim#1{\mathop{\vtop{\ialign{##\crcr
     $\hfil\displaystyle{#1}\hfil$\crcr\noalign
     {\kern1pt\nointerlineskip}\hbox{$\hfil\sim\hfil$}\crcr
     \noalign{\kern1pt}}}}}

\def\eg{{\it e.g.\ }}  \def\ie{{\it i.e.\ }}
\def\cf{{\it cf.\ }}

\usepackage{amsmath,amssymb, bbm} 
\usepackage{graphicx}

\newcommand{\ncd}{\newcommand} 
\ncd{\nms}{\negmedspace} 
\ncd{\nts}{\negthickspace} 
\ncd{\mcl}[1]{\mathcal{#1}} 
\ncd{\beq} {\begin{equation}} 
\ncd{\eeq} {\end{equation}} 
\ncd{\BE} {\begin{eqnarray}} 
\ncd{\EE} {\end{eqnarray}} 
\ncd{\rarr} {\rightarrow} 
\ncd{\larr} {\leftarrow} 
\ncd{\lrarr} {\leftrightarrow} 
\ncd{\lbeq}[1]  {\label{eq: #1}} 
\ncd{\refeq}[1] {(\ref{eq: #1})} 
\ncd{\mrm}    {\mathrm} 
\ncd{\nn}{\nonumber} 
\ncd{\mbf}[1] {{\mathbf #1}} 
\ncd\T{\frac{1}{2}h^{\mu\nu}p_\mu p_\nu} 
\ncd{\ms}{\mathstyle} 
\ncd{\ds}{\displaystyle} 
\ncd{\bmth}[1] {\mbox{\boldmath $#1$}} 
\ncd{\abs}[1] {|#1|} 
\ncd{\ubold}{\mathbf u}
\ncd{\Abold}{\mathbf A}
\ncd{\Bbold}{\mathbf B}
\ncd{\Mbold}{\mathbf M}
\ncd{\tsfrac}[2]{{\ts\frac{#1}{#2}}}
\ncd{\lagom}{\hspace{.6pt}}
\ncd{\muk}{k}
\ncd{\dumkonstant}{v_0}

\newtoks\reportnoregister \newtoks\eprintnoregister
\newcommand{\reportnumber}[1]{\reportnoregister={#1}}
\newcommand{\eprintnumber}[1]{\eprintnoregister={#1}}

\reportnumber{\mbox{}} 
\eprintnumber{\mbox{}} 

\newcommand{\reportid}{
   \begin{minipage}{17cm}\vspace{-7.2cm}
     \begin{flushright}
      {\normalsize \the\reportnoregister \\[-.2cm]
            \eprint{\the\eprintnoregister}}\vspace{0.2cm}
     \end{flushright}
   \end{minipage}\hspace{-17cm} }

\catcode`@=11   
\def\title#1{\gdef\@title{\reportid#1}}
\catcode`@=12   

\ncd{\gtnoll}{{}^{0\!}\eta}
\ncd{\gnoll}{{}^{0\!}\eta}
\ncd{\bgtnoll}{{}^{0\!}\pmb{\eta}}

\begin{document} 

\reportnumber{USITP 2002-5}
\eprintnumber{gr-qc/0211026}

\title{Elastic Stars in General Relativity I \\ Foundations and Equilibrium Models}
  \author{Max Karlovini\footnote{E-mail: \eprint{max@physto.se}} \:and
Lars Samuelsson\footnote{E-mail: \eprint{larsam@physto.se}} \\[10pt]
  {\small Department of Physics, Stockholm University}  \\
  {\small SCFAB, 106 91 Stockholm, Sweden} }
\date{}
\maketitle


\begin{abstract}{\normalsize
  The aim of this paper is twofold. First, we set up the theory of
  elastic matter sources within the framework of general relativity in
  a self-contained manner. The discussion is primarily based on the
  presentation of Carter and Quintana but also includes new methods
  and results as well as some modifications that in our opinion make
  the theory more modern and transparent.  For instance, the equations
  of motion for the matter are shown to take a neat form when
  expressed in terms of the relativistic Hadamard elasticity
  tensor. Using this formulation we obtain simple formulae for the
  speeds of elastic wave propagation along eigendirections of the
  pressure tensor. Secondly, we apply the theory to static spherically
  symmetric configurations using a specific equation of state and
  consider models either having an elastic crust or core.}
\end{abstract}
\vspace{.5cm}
\centerline{\bigskip\noindent PACS: 04.40.Dg, 26.60.+c, 97.10.Cv, 97.60.Jd}

\section{Introduction}

The presence of a solid crust in a neutron star is well established on
theoretical grounds\cite{pr:nscrust}. There are also speculations of
solid cores due to the dominant strong interaction in the deep
interiors of such stars. Although the presence of a solid crust is not
thought to influence the mass or radius of a neutron star
substantially there are still a number of other observational issues
attributable to such a crust. The best observational data of what goes
on inside a neutron star crust comes from pulsar timing measurements,
occasionally showing sudden spin-ups in the otherwise very regular
spin-down pattern. These events, called glitches, are thought to be
due to interactions between the crust and superfluid neutrons
penetrating the lattice see \eg \cite{sauls:superfluid}. Another
interesting feature observed in some pulsars is a smooth modulation
of the timing measurements. This could be caused by free precession of
a non-axisymmetric configuration of the crust with respect to the
angular momentum axis. For discussions on this phenomenon see \eg
\cite{ja:precess, cul:precession}.  Of course, a crust has many other
interesting implications including for instance its effects on the
(quasi) normal modes of the star. For a nonrotating newtonian perfect
fluid star all axial modes are trivial in the sense that they are
degenerate at zero frequency and hence correspond to non-oscillatory
flows. The presence of the crust breaks the zero-frequency degeneracy
of the axial modes as well as adding new modes due to the interface to
the fluid core. For calculations and discussion within the Newtonian
framework see \cite{mvh:osc, strohetal:shearmodulus} and references
therein. Axial modes of relativistic initially isotropic elastic stars
have also been considered in some detail in \cite{st:torsional}. For
some very recent relativistic calculations within the Cowling
approximation and using a Hookean equation of state see
\cite{yl:nonrad}. 

So far, all studies have been based on some kind of weak field
approximation. Either it has been assumed that the gravitational field
is weak or not coupled to matter oscillations, or that the deviation
from an unsheared material state is small and can be treated linearly
by a Hookean approximation. To be able to study nonlinear effects that
fall outside such approximation schemes it is desirable to develop the
general relativistic theory of elasticity to the point where it can be
used in practical calculations. We intend for this work to be a step
towards that goal.

Although classical newtonian elasticity theory has a long history,
going back to the 17th century and Hooke's law, the development of a
theory of elasticity adapted to general relativity took off rather
late. The first demand for such a theory came in the late 50s with
Weber's bar antenna for gravitational waves\cite{weber:bar},
necessitating an understanding of how these waves interact with
elastic solids. For this application it is sufficient to use the
linearised version of Einstein's equation with a background that can
be treated newtonianly.  Indeed, this week-field approximation was
used by Weber and was further developed in subsequent works by
Dyson\cite{dyson:seismic} and others. Following
unsuccessful\cite{synge:elastic}, partially
successful\cite{rayner:elastic} and successful yet rather
coordinate-cluttered\cite{oldroyd:contmatter} attempts, a fully
developed non-linear theory of elasticity adapted to general
relativity was given in 1973 in a paper by Carter and
Quintana\cite{cq:elastica} (hereinafter referred to as CQ), which
still is a standard reference in the field. However, as later noted by
Carter\cite{carter:rheometric}, the basic theoretical framework of
their theory had already been given by Souriau\cite{souriau:elastic},
a work that largely had passed unnoticed. Also preceding CQ was Maugin
who has made substantial contributions to the field, including the
development of a theory of polarizable elastic
media\cite{maugin:magnetized,maugin:elastic}, which should be
important for realistic modeling of neutron stars since they are
believed to have strong magnetic fields. Carter also gave a
discussion of such matter in \cite{carter:rheometric}. In more recent
years, the general relativistic elasticity theory has been
reconsidered by Magli and
Kijowski\cite{mk:elastolagrangian,mk:elastohamiltonian} who put
emphasis on the gauge character of the theory, and by
Christodoulou\cite{chris:crystalcontinua}. Very recently a number of
existence and uniqueness theorems have been proved by Beig and
Schmidt\cite{bs:relasticity}.

We would like to emphasise that the above bibliographical remarks by
no means give a complete history of the subject of general
relativistic elasticity theory. For further references see for
instance \cite{mk:elastolagrangian, mk:gaugetype}.

In this paper we set up the theory of general relativistic elasticity
in a self-contained manner taking as our starting point the theory
described in CQ. Although our treatment follows CQ rather closely and
sometimes in detail, new methods and results are presented along the
way. We do not want the reader to believe that we consider CQ to be
superior to other sound presentations of the theory in any fundamental
way. We simply choose it as a standard reference to avoid introducing
yet another completely new set of notations and definitions into a
field which already suffers from being a bit of a jungle when it comes
to comparing results from different workers. Also, being relativists,
we favor a formulation of the theory in terms of pure spacetime
tensor equations, rather than formulations that keep material space
indices and coordinates in the formulae.

The theoretical part of the paper is followed by an application
to static spherically symmetric configurations. Having a thorough
control and understanding of such models is essential as they provide
the backgrounds for more elaborate applications, \eg slowly rotating
stars via the Hartle-Thorne formalism\cite{hartle:slowrot, ht:slowrot}
or stellar oscillations of various kinds. Spherically symmetric
elastic matter models appears to have been first studied by Magli and
Kijowski\cite{mk:nonrot} and were later also investigated by
Park\cite{park:elastsss}.  However, in this paper our aim is to give a
recipe for numerically producing physically relevant prestressed
stellar models with speeds of wave propagation that are explicitly
known. Such models will serve as backgrounds in subsequent
investigations of stellar perturbations.  The paper is organised as
follows:

In section \ref{sec:relasticity} we introduce and discuss the general
concepts of the theory, based on the foundation that the matter field
is a mapping from the four-dimensional spacetime manifold to an
abstract three-dimensional manifold. The latter will be referred to as
the material space, since its points may be thought of as the material
particles. To insure that the theory is invariant under coordinate
transformations on both manifolds, the Lagrangian of the theory, which
is directly given by the energy density, will be a function of scalars
that can be formed by contractions of the inverse spacetime metric
with covariant tensors that are pulled back from the material space.
We feel that this is the most natural way to make the coordinate
invariance inherent to the theory, as well as to have a clear
distinction between the spacetime metric and the material space
mapping as the gravitational and matter field variables of Einstein's
equations with elastic source. This approach is slightly different
from the one used in CQ, where the energy density is viewed as a
function of the spatial part of the spacetime metric (which depends
both on the spacetime metric itself as well as on the matter field),
implicitly or explicitly contracted with contravariant tensors that
can be identified as material space tensors. Using the spatial metric
as a ``variable'' in this way would make it cumbersome to fix the
spacetime metric and only consider variations of the matter field,
which has been a source of critique from other workers (\cf the
appendix of \cite{maugin:prestressed} and also
\cite{mk:gaugetype}). However the critique is partially 
unjustified since it seems to be based on the assumption that the
spatial metric is to be viewed as the elastic matter field.

Furthermore, the basic mathematical tools that we use are standard
tools of differential geometry such as the pull-back and push-forward
associated with the material space mapping. This means that we are
using a somewhat lighter mathematical machinery than CQ who are
explicitly making use of the isomorphism of the spacetime tangent
subspaces orthogonal to the material flowlines and the tangent spaces
of the material space manifold, as well as an extended version of the
convected derivative first introduced by
Oldroyd\cite{oldroyd:contmatter}. The reason that we are not
discussing or using these concepts is not because we by any means feel
that they are irrelevant, but because we, during the progress of this
work, found that they were not needed for any of our purposes. For a
thorough treatment of them, we instead refer the reader to CQ. The
important results of this section is the general form of the
stress-energy tensor and its specialization to the case when a fixed
metric is the only material space tensor field which is used to set up
the equation of state, an assumption which will be made throughout the
paper unless explicitly stated otherwise.

In section \ref{sec:eom} we discuss the matter equations of motion,
\ie the divergence-free condition for the stress-energy tensor. To
this end we introduce a spatially projected connection on spacetime
which can be viewed as the ``pull-back'' of the Levi-Cevita connection
associated with the material space metric. This projected connection is
given by a three index tensor field which we call the relativistic
elasticity difference tensor. We then show that the Euler equations,
to which the matter equations of motion reduces, can be put in an
elegant form involving the four index relativistic Hadamard elasticity
tensor and the relativistic elasticity difference tensor.  To our
knowledge, this form of the Euler equations is first given here. We
also discuss how to generalise this method to more general equations
of state involving material space tensor fields other than a metric.

In section \ref{sec:eig} we derive various useful relations that
involve the eigenvectors and eigenvalues of the pull-back of the
material space metric. This leads us to introduce three scalars that
we refer to as the linear particle densities which simply are defined
as the square roots of the eigenvalues. The advantage of working with
these linear particle densities is that we arrive at several formulae
that closely resemble analogous formulae for perfect fluids when the
latter are described in terms of the (volume) particle density. 
%

In section \ref{sec:sound} we derive the characteristic equation that
determines the speed of wave propagation in relativistic elastic
media. We do so by employing the standard Hadamard method of
considering field discontinuities across a hypersurface, in a manner
similar to Carter\cite{carter:sound}.  However, by starting out from
our form of the nonlinear equations of motion we obtain a more direct
link between the full nonlinear theory and the theory of wave
propagation. The characteristic equation is explicitly solved in the
case when the wave propagates in a principal (\ie eigenvector)
direction and we obtain a strikingly simple formula for the speed of
the longitudinal waves, closely analogous to the formula for the speed
of sound in perfect fluids. For the transversal waves, which have no
analogue for perfect fluids, the propagation speed is found to be
given by a simple algebraic formula involving the principal pressures
and the linear particle densities. Although wave propagation in
prestressed nonlinear elastic solids has been treated relativistically
also by Maugin in a number of
papers\cite{maugin:prestressed,maugin:discont,maugin:crust,maugin:singular},
we feel that our approach and usage of the linear particle densities
make the derivation as well as the resulting formulae much more
transparent.

In section \ref{sec:eos} we introduce a particular class of equations
of state which much resemble the quasi-hookean type of equations of
state described in CQ, the sole difference being that we use a
different definition of the shear scalar which we feel is more
convenient. Explicit formulae for the principal pressures and speeds
of sound propagation are given.

Section \ref{sec:deg} is concerned with the degenerate case when two
of the linear particle densities coincide. The reason why we pay
particular attention to this nongeneric case is that any spherically
symmetric configuration by necessity belongs to it. 
 
In section \ref{sec:sss} we set up Einstein's equations for static
spherically symmetric configurations with an elastic matter source.
The full set of equations of motion are found to form a
(nonautonomous) system of three first order ordinary differential
equations. This should be contrasted to the perfect fluid case for
which the number of equations is two.  

In section \ref{sec:bc} we discuss the boundary conditions that static
spherically symmetric configurations must satisfy. In particular we
show that transitions between a material phase interior to an elastic
solid form a one-parameter set of phase transitions parametrizable by
the discontinuity in the tangential pressure. Physically speaking, the
nature of such a phase transition is determined by the interaction
between interior and exterior phases as well as the history of the
configuration. However it is not directly given by the matching
conditions of the Einstein equations. This is in contrast to the
perfect fluid case where the phase transitions are given by the
equation of state in a direct way.

We then proceed in section \ref{sec:model} to integrate some specific
toy models of neutron stars using a relativistic polytropic equation
of state as description of the unsheared behaviour of the matter.  We
let the shear modulus be proportional to the unsheared pressure and
study the effects of on the one hand including a solid core, and on
the other having a rigid crust. As expected the equilibrium models do
not differ very much from the corresponding perfect fluid ones for
realistic values of the shear modulus.

In the Appendix we discuss and motivate the class of equations of
state introduced in section \ref{sec:eos}, including our unorthodox
shear scalar definition. The discussion is based on a second order
expansion of a general equation of state around an arbitrarily chosen
unsheared state.

A few words on the notation and conventions used in this paper should
be mentioned here. Tensors will be given in the abstract index
notation in the manner described in Wald\cite{wald:gr}. Abstract
spacetime indices will be denoted by lowercase Latin letters
$(a,b,\ldots)$, whereas material space indices will be denoted by
capital Latin letters $(A,B,\ldots)$. Greek letters $(\mu, \nu,
\ldots)$ or $(\Lambda,\Sigma,\ldots)$ will be used as indices 
referring to a given basis and for these the Einstein summation
convention will not be used. In order to avoid cluttering formulae by
explicitly writing out arguments we will use the convention that
quantities depending only on the particle density will be denoted by a
check accent. For example, the symbol $\check{p}$ will denote
unsheared pressure, to be explained subsequently. The metric is taken
to have signature +2 and the units are such that the speed of light is
equal to unity and the Einstein equations take the form $G_{ab}=\kappa
T_{ab}$ with $\kappa$ being given in arbitrary units.

\section{Relasticity}\label{sec:relasticity}
The theory of elastic matter sources in general relativity that we
will use is based on the presentation of Carter and
Quintana\cite{cq:elastica}. Since we expect that this theory is not
generally known, we will start the discussion by outlining the basic
ideas. To begin with, we introduce an abstract three-dimensional
manifold, the \emph{material space}, $X$, whose points are to be
thought of as the material particles when going to the continuum
limit. This space can in general be equipped with various types of
tensor fields which give the structure of the material in a reference
state. To begin with, the only tensor field that we introduce is,
using capital Latin tensor indices on $X$, a \emph{particle density
  form} $n_{ABC}=n_{[ABC]}$. This three-form is defined in such a way
that when integrated over some volume in $X$, the result is simply the
number of particles contained in that volume.

A \emph{state of matter} in a material filled open four-dimensional
submanifold $M'\subseteq M$, with $M$ being the full spacetime
manifold, is given by the spacetime metric $g_{ab}$ restricted to $M'$
and a $C^1$ (at least) mapping
\begin{equation}
  \psi : M' \rarr X.
\end{equation}
This mapping $\psi$ should be thought of as the matter field of the
theory, analogous to a scalar field but with the difference that
$\psi$ maps to a three-dimensional space while the scalar field maps
to a one-dimensional one. We shall always assume that the preimage
$\psi^{-1}(p)$ of all points $p\in X'=\psi(M')$ is a single timelike
curve in $M'$, which is to be interpreted as the flowline of the
particle represented by the point $p$. This way $\psi$ gives rise to
an identification of the three-dimensional manifold of particle
flowlines with the material space submanifold $X'$. The push-forward
$\psi_*$ and pull-back $\psi^*$, taking contravariant tensors from
$M'$ to $X'$ respectively covariant tensors from $X'$ to $M'$, are
important ingredients when setting up the theory.  We shall use the
convention that the push-forward of a contravariant spacetime tensor
$t^{a\ldots}$ will be denoted by $t^{A\ldots}$ and similarly that the
pull-back of a covariant material space tensor $t_{A\ldots}$ will be
denoted by $t_{a\ldots}$. In other words we simply replace the lower
case Latin spacetime indices with capital Latin material space indices
and vice versa. From the made assumptions about $\psi$ it follows that
the kernel of $\psi_*$, when viewed as a linear mapping of spacetime
vectors $v^a$ to material space vectors $v^A$, is comprised of the
vectors that are tangent to the particle flowlines.  It can be proved,
using the commutation relations involving pull-back, push-forward,
tensor contraction and Lie differentiation, that given a covariant
spacetime tensor field $t_{a\ldots}$, there exists a fixed material
space tensor field $t_{A\ldots}$ whose pullback is $t_{a\ldots}$ if
and only if the following conditions hold for all vector fields $v^a$
that are everywhere flowline tangential;
\begin{itemize}
\item{All contractions of $v^a$ with any index of $t_{a\ldots}$
    vanishes, \ie   $t_{a\ldots}$ is completely flowline orthogonal.}
\item{$t_{a\ldots}$ is Lie dragged by $v^a$; $\mathcal{L}_v
    t_{a\ldots}$ = 0}.
\end{itemize}
The pulled back particle density form $n_{abc} = \psi^*\,n_{ABC}$ is
of fundamental importance for the theory and is used to define the
flowline tangential \emph{particle current}
\begin{equation}\lbeq{ncurrentdef}
  n^a = \tsfrac1{3!}\,\epsilon^{abcd}n_{bcd},
\end{equation}
where $\epsilon_{abcd}$ is the spacetime volume form associated with
$g_{ab}$. By construction, $n^a$ is conserved, 
\begin{equation}
  \nabla_{\!a}n^a = 0,
\end{equation}
which follows from the fact that exterior differentiation commutes
with the pull-back operation, implying that $n_{abc}$ is a closed
spacetime three-form due to $n_{ABC}$ trivially being a closed
material space three-form. Noting that $n^a$ by assumption is
timelike, we set 
\begin{equation}
  n^a = nu^a, \quad n = \sqrt{-n^a n_a}, \quad u^a u_a = -1,
\end{equation}
where $n$ is the \emph{particle density} and $u^a$ the matter
four-velocity. To proceed we introduce the spatial volume form
\begin{equation}
  \epsilon_{abc} = \epsilon_{abcd}u^d. 
\end{equation}
From eq.\ \refeq{ncurrentdef} it follows that 
\begin{equation}\lbeq{nepsilon}
  n_{abc} = n\,\epsilon_{abc},
\end{equation}
which justifies the interpretation of $n$ as the particle density. 

An \emph{equation of state} is supplied by giving the rest frame
energy density $\rho$ as a function of scalars that are formed by
contracting the inverse metric $g^{ab}$ with pulled back covariant
material space tensor fields. The particle density may be used as one
of the scalars, as indeed, according to eq.\ \refeq{nepsilon}, it is
given by the contraction
\begin{equation}\lbeq{ncontract}
  n^2 = \tsfrac1{3!}\,n^{abc}n_{abc} =
\tsfrac1{3!}\,g^{ad}g^{be}g^{cf}n_{abc}n_{def}.  
\end{equation}
Once an equation of state is given, the stress-energy tensor
takes the form
\begin{equation}\lbeq{stressenergy}
T_{ab} = -\rho g_{ab} + 2\frac{\partial\rho}{\partial g^{ab}} =
\rho\,u_a u_b + p_{ab},  
\end{equation}
where
\begin{equation}\lbeq{stress}
  p_{ab} = 2\frac{\partial\rho}{\partial g^{ab}} - \rho h_{ab}, \quad
 u^a p_{ab} = 0,
\end{equation}
and
\begin{equation}
  h_{ab} = u_a u_b + g_{ab}.
\end{equation}
The above equations can be taken as the definition of the
stress-energy tensor for elastic matter, but it is worth noting that
$T_{ab}$ can also be straightforwardly derived 
according to
\begin{equation}
  T_{ab} = -\frac2{\sqrt{-g}}\frac{\delta S_M}{\delta g^{ab}}
\end{equation}
from the matter action
\begin{equation}\lbeq{SM}
  S_M = -\int dx^4\sqrt{-g}\rho. 
\end{equation}
It will prove convenient to rewrite the energy density as $\rho =
n\epsilon$ where $\epsilon$ is the energy per particle.  For instance
the pressure tensor $p_{ab}$ is then given by the simple formula
\begin{equation}\lbeq{pressuretensor}
p_{ab} = 2n\frac{\partial\epsilon}{\partial g^{ab}},    
\end{equation}
where we have used that
\begin{equation}\lbeq{dndh}
  \frac{\partial n}{\partial g^{ab}} = \tsfrac12 n h_{ab},
\end{equation}
which follows, for instance, from eq.\ \refeq{ncontract}.
Now, scalars that are formed on spacetime by contracting $g^{ab}$ with
pulled back tensors $t_{a\ldots} = \psi^*t_{A\ldots}$ can
alternatively be formed on the material space by making the analogous
contractions of the pushed forward tensor $g^{AB} = \psi_* g^{ab} = 
\psi_* h^{ab}$ with the tensors $t_{A\ldots}$. Hence, at all points
in $X'$, we may view the energy density $\rho$ as a function of
$g^{AB}$ and fixed material space tensors $t_{A\ldots}$. Taking this
view we shall make the assumtion that $\epsilon$ has a minimum value
$\check{\epsilon}$ under variations of $g^{AB}$ such that the particle
density $n$ is held fixed. Such a state will be referred to as an
\emph{unsheared} state. This allows us to introduce an $n$-dependent
tensor $\eta_{AB}$ on the material space such that
$g^{AC}\eta_{\,CB}=\delta^A{}_B$ when $\epsilon=\check\epsilon$. In
other words, we define $\eta_{AB}$ such that the minimum of
$\epsilon(g^{AB})$ occurs at $g^{AB}=\eta^{-1AB}$. If
$\check\epsilon$, which is a function of $n$ only, has a minimum at a
certain particle density $n=n_0$, it follows that $\epsilon$ has an
absolute minimum under \emph{all} variations of $g^{AB}$. This
minimizing state is referred to as a \emph{completely unstrained} or
\emph{completely relaxed} state. We shall not assume the existence of
such a state simply because the material structure in the interiors of
a neutron star may owe its existence to the high pressures there.

It is fairly straightforward to show that the pull-back of the
Levi-Cevita volume form of $\eta_{AB}$ coincides with the spatial
volume form $\epsilon_{abc}$. Hence we let the volume form of
$\eta_{AB}$ be denoted $\epsilon_{ABC}$ as we have
$\psi^*\epsilon_{ABC} = \epsilon_{abc}$. Clearly, the relation between
$\epsilon_{ABC}$ and the particle density form $n_{ABC}$ is analogous
to eq.\ \refeq{nepsilon} for their pull-backs $\epsilon_{abc}$ and
$n_{abc}$, \ie
\begin{equation}\lbeq{volform}
  n_{ABC} = n\,\epsilon_{ABC}.
\end{equation}
It is worth keeping in mind that, despite the notation, the particle
density form $n_{ABC}$ is in fact a fixed material space tensor,
independent of $n$. It is convenient to define a new tensor $k_{AB}$,
conformal to $\eta_{AB}$ and having $n_{ABC}$ as its volume form. It
follows from eq.\ \refeq{volform} that this tensor is given by
\begin{equation}\lbeq{kdef}
  k_{AB}=n^{2/3}\eta_{AB}
\end{equation}
Denoting differentiation with respect to $n$ by a prime we have
\begin{equation}\lbeq{deta}
  n\,\eta\,'_{\!AB}=-\tsfrac23\eta_{AB}+\tau_{AB},
\end{equation}
where 
\begin{equation}\lbeq{taudef}
  \tau_{AB}=n^{1/3}k\,'_{\!AB}
\end{equation}
Now, since $k^{-1AB}k\,'_{\!AB}=0$ due to the fact that the
determinant of $k_{AB}$ is independent of $n$ (since $n_{ABC}$ is), we
find that the trace of eq.\ \refeq{deta} with respect to
$\eta^{-1AB}=n^{2/3}k^{-1AB}$ is
\begin{equation}
  n\,\eta^{-1AB}\eta\,'_{\!AB}=-2
\end{equation}
Thus $\tau_{AB}$, called the \emph{compressional distortion tensor},
is clearly the trace-free part of $n\,\eta\,'_{\!AB}$ and therefore
measures how the deformation of the unsheared crystal structure
deviates from pure conformal rescaling when varying the particle
density. It now follows from the definition \refeq{taudef} that
$k_{AB}$ is independent of $n$ precisely when the material deforms
conformally. In this paper we will mainly consider such materials, which
include materials with isotropic but also, for instance, cubic
unsheared structures. Consequently, we will have a fixed
$n$-independent metric tensor field $k_{AB}$ on $X$. We shall also
make a further restricting assumption, namely that the energy per
particle $\epsilon$ is a function only of the independent invariants
that can be formed from the flowline orthogonal mixed tensor
\begin{equation}
  k^a{}_b = g^{ac}k_{cb}, \quad k_{ab} = \psi^*k_{AB}.
\end{equation}
When acting with $\partial/\partial g^{ab}$ on quantities that depend
only on $k^a{}_b$ we may reexpress this operator as
\begin{equation}\lbeq{dellhab}
 \frac{\partial}{\partial g^{ab}} = k_{c(a}\frac{\partial}{\partial
 k^{b)}{}_c},
\end{equation}
which follows from 
\begin{equation}\lbeq{dkdh}
 \frac{\partial k^c{}_d}{\partial g^{ab}} = \delta^c{}_{(a}k_{b)d}.
\end{equation}
Since $k^a{}_b$ is restricted to be orthogonal to $u^a$, it has three
independent scalar invariants, which for instance could be taken to be
\begin{equation}
  I_1 = k^a{}_a, \quad I_2 = k^a{}_b k^b{}_a, \quad I_3 =k^a{}_b
  k^b{}_c k^c{}_a.
\end{equation}
With $n_{ABC}$ being the volume form of $k_{AB}$, it follows that the
particle density $n$ is in fact also a scalar invariant of $k^a{}_b$,
given by
\begin{equation}\lbeq{ndet}
  n^2 = \det{(k^a{}_b)} = \tsfrac1{3!}\,(I_1^{\,3} - 3 I_1 I_2 + 2I_3). 
\end{equation}
Rather than using $\{I_1,I_2,I_3\}$ as the set of three independent
scalar invariants of $k^a{}_b$, it is both convenient and instructive
to instead take the set to be comprised of $n$ and two independent
scalar invariants of the mixed tensor
\begin{equation}
  \eta^a{}_b = h^{ac}\eta_{cb}, \quad \eta_{ab} = \psi^*\eta_{AB},
\end{equation}
whose deviation from $h^a{}_b$ gives all information about
deformations other than conformal compressions. From the relation
$k^a{}_b = n^{2/3}\eta^a{}_b$ it is clear that $\det{(\eta^a{}_b)} =
1$, whence $\eta^a{}_b$ has two independent invariants rather than
three, due to its invariance under conformal compressions $h_{ab}\rarr
C^{-2}h_{ab}$ under which $k^a{}_b\rarr C^2 k^a{}_b$ and $n\rarr
C^3n$. We shall now obtain an alternative expression for the pressure
tensor by using that the decomposition $k^a{}_b = n^{2/3}\eta^a{}_b$
of $k^a{}_b$ into a conformal factor $n^{2/3}$ times a unit
determinant mixed tensor $\eta^a{}_b$ gives rise to the following
splitting of $\partial/\partial g^{ab}$ valid when acting on
invariants (scalar or tensorial) of $k ^a{}_b$:
\begin{equation}\lbeq{ddhneta}
 \frac{\partial}{\partial g^{ab}} = \tsfrac12
 nh_{ab}\frac{\partial}{\partial n}+
 \eta_{c<a}\frac{\partial}{\partial \eta^{b>}{}_c}, 
\end{equation}
where the angle brackets $<\,\,,\,>$ around an index pair here denotes
the symmetric and traceless part of that pair, where ``traceless
part'' refers to $h_{ab}$ and not the full spacetime metric $g_{ab}$.
Explicitly, for a tensor $t_{ab}$, we have
\begin{equation}
  t_{<ab>} = t_{(ab)} - \tsfrac13 t^c{}_c h_{ab}.
\end{equation}
In order to obtain eq.\ \refeq{ddhneta} we have used eq.\ \refeq{dndh} and
\begin{equation}\lbeq{dgammatildedh}
  \frac{\partial\eta^c{}_d}{\partial g^{ab}} = h^c{}_{<a}\eta_{\,b>d}
\end{equation}
which follows from $\eta^c{}_d = n^{-2/3}k^c{}_d$ and eqs.\ 
\refeq{dndh} and \refeq{dkdh}. 
So, by specifying $\epsilon$ as a function of $n$ and
$\eta^a{}_b$, the pressure tensor can be written as
\begin{equation}\lbeq{SET}
  p_{ab} = p\,h_{ab} + \pi_{ab}, \quad \pi^a{}_a = 0,
\end{equation}
where
\begin{align}
  &p = n^2\frac{\partial\epsilon}{\partial n}, \\ \lbeq{pieta}
  &\pi_{ab} = 2n\,\eta_{c<a}\frac{\partial\epsilon}{\partial
  \eta^{\,b>}{}_c}. 
\end{align}
This clearly displays that the dependence of $\epsilon$ on the
particle density $n$ is directly related to the \emph{isotropic
  pressure} $p$, while the dependence on the unimodular tensor
$\eta^a{}_b$ is directly related to the trace-free \emph{anisotropy
  pressure tensor} $\pi_{ab}$. In particular, we see that the matter
source can be interpreted as a perfect fluid iff the energy density is
a function of the number density alone, the equation of state then
being $\rho = n\epsilon$, $p=n^2 d\epsilon/dn$ where $\epsilon$
coincides with $\check\epsilon$, the minimal energy per particle at
fixed particle density.
\section{Equations of motion for elastic matter}
\label{sec:eom}

Let us now turn our attention to the equations of motion $\nabla_b
T^{ab}=0$, which can also be obtained by varying the matter action
\refeq{SM} with respect to the material space mapping $\psi$. For 
any stress-energy tensor of the form
\begin{equation}
  T_{ab} = \rho u_au_b + p_{ab}, \quad u^au_a=-1, \quad u^ap_{ab}=0,
\end{equation} 
the divergence free condition splits into its projection onto and
orthogonal to $u^a$. In terms of the spatially projected connection
$D_a$ defined by its action on an arbitrary tensor
$t^{b\ldots}{}_{c\ldots}$ according to
\begin{equation}
  D_a t^{b\ldots}{}_{c\ldots} = h^d{}_ah^b{}_e\cdots h^f{}_c\cdots
  \nabla_d t^{e\ldots}{}_{f\ldots},
\end{equation}
the equations of motion takes the form
\begin{align}
  \dot\rho+(\rho h^{ab}+p^{ab})\Theta_{ab} &=0, \lbeq{rhodot}\\
  (\rho h^{ab}+p^{ab})\dot{u}_b+D_bp^{ab} &= 0, \lbeq{divp}
\end{align}
where, as usual, a dot denotes covariant derivative along $u^a$ and
\begin{equation}
  \Theta_{ab} = D_{(a}u_{b)}. 
\end{equation}
For elastic matter, when the stress-energy tensor is given by an
equation of state according to eqs.\ \refeq{stressenergy} and
\refeq{stress}, it follows that eq.\ \refeq{rhodot} is in fact
automatically satisfied, leaving the $u^a$ orthogonal \emph{Euler
equations} as the only nontrivial matter equations of motions.

When $u^a$ is hypersurface orthogonal the projection tensor $h_{ab}$
is the metric induced on these hypersurfaces. In this case the
operator $D_a$ is the Levi-Cevita connection associated with $h_{ab}$.
In general, however, $D_a$ cannot be viewed as a connection defined on
a family of spacetime submanifolds, but it can nevertheless be viewed
as the Levi-Cevita ``pseudo-connection'' of $h_{ab}$ in the sense that
the property
\begin{equation}
\lbeq{dhab}
  D_a h_{bc} = 0
\end{equation}
always holds.  Since we are assuming that the material space is
equipped with a fixed metric $k_{AB}$, its pullback $k_{ab}$ provides
an alternative metric on the tangent subspaces orthogonal to $u^a$.
Furthermore, we may also carry over the Levi-Cevita connection
$\tilde{D}_A$ of $k_{AB}$ into a differential operator $\tilde{D}_a$
acting on spacetime tensors. To do so we declare that $\tilde{D}_a$,
like $D_a$, should be a projected torsion-free connection with respect
to the projection tensor $h^a{}_b$, \ie there should exist a
torsion-free connection $\tilde\nabla_a$ on $M$ such that the action
of $\tilde{D}_a$ on an arbitrary tensor field
$t^{b\ldots}{}_{c\ldots}$ is given by
\begin{equation}\lbeq{Dtildet}
  \tilde{D}_a t^{b\ldots}{}_{c\ldots} = h^d{}_ah^b{}_e\cdots h^f{}_c\cdots
  \tilde\nabla_d t^{e\ldots}{}_{f\ldots}.
\end{equation}
It follows that $\tilde{D}_a$ is uniquely determined by $D_a$ and a
completely flowline orthogonal tensor field $C^c{}_{ab}$ such that for
all spacetime vector fields $X^c$, we have 
\begin{equation}\lbeq{Cintro}
  \tilde{D}_a X^c = D_aX^c + C^c{}_{ab}X^b. 
\end{equation}
We will refer to the tensor field $C^c{}_{ab}$ as the relativistic
\emph{elasticity difference tensor}, since 
the tensor field that gives the difference between two connections 
is sometimes called the difference
tensor in the mathematical literature\cite{spivak:diffgeom2}. To relate
$\tilde{D}_a$ to $\tilde{D}_A$ we impose that for any spacetime vector
$X^a$ and vector field $Y^a$, it should hold true that
\begin{equation}\lbeq{Dtildeimpose}
  \psi_*(X^b\tilde{D}_b Y^a) = X^B\tilde{D}_B Y^A, \quad X^B =
  \psi_*X^b, \quad Y^A = \psi_*Y^a.
\end{equation}
This fixes $C^c{}_{ab}$ to be the tensor field
\begin{equation}
  C^c{}_{ab} = \tsfrac12 k^{-1\,cd}(D_a k_{bd} + D_b k_{ad} - D_d
  k_{ab}),
\end{equation}
with $k^{-1\,cd}$ being the flowline orthogonal inverse of $k_{ab}$,
\begin{equation}
  k^{-1\,ac}k_{cb} = h^a{}_b, \quad k^{-1\,ab}u_b = k^{-1\,ba}u_b =
  0. 
\end{equation}
Clearly, the tensor $C^c{}_{ab}$ has the symmetry
\begin{equation}\lbeq{Csymm}
  C^c{}_{ab} = C^c{}_{(ab)}.
\end{equation}
Moreover, as one might guess, the projected connection $\tilde{D}_a$
satisfies
\begin{align}\lbeq{Dtildepushfwd}
  &\psi_*(X^b\tilde{D}_b t^{a\ldots}) = X^B\tilde{D}_B t^{A\ldots},
  \quad X^B = \psi_*X^b, \quad t^{A\ldots} = \psi_*t^{a\ldots} \\ \lbeq{Dtildepullback}
  &\psi^*(\tilde{D}_B t_{A\ldots}) = \tilde{D}_b t_{a\ldots}, \quad 
  t_{a\ldots} = \psi^* t_{A\ldots}.
\end{align}
In particular, eq. \refeq{Dtildepullback} implies
\begin{equation}\lbeq{tildeDk}
  \tilde{D}_c k_{ab} = 0.
\end{equation}
Letting $t^{b\ldots}{}_{c\ldots}$ be a tensor field built up out of
$g^{ab}$ and $k_{ab}$, it follows from eq.\ \refeq{tildeDk} that 
\begin{equation}
  \tilde{D}_a t^{b\ldots}{}_{c\ldots} = \frac{\partial
  t^{b\ldots}{}_{c\ldots}}{\partial g^{de}}\tilde{D}_a g^{de} =
  \frac{\partial t^{b\ldots}{}_{c\ldots}}{\partial
  g^{de}}(\tilde{D}_a-D_a)g^{de} = 2\frac{\partial
  t^{b\ldots}{}_{c\ldots}}{\partial g^{de}}\,C^{de}{}_a,
\end{equation}
implying in turn that
\begin{equation}
  D_a t^{b\ldots}{}_{c\ldots} = \tilde{D}_a t^{b\ldots}{}_{c\ldots} -
  (\tilde{D}_a-D_a)t^{b\ldots}{}_{c\ldots} = 2\frac{\partial
  t^{b\ldots}{}_{c\ldots}}{\partial g^{de}}C^{de}{}_a -
  t^{d\ldots}{}_{c\ldots}C^b{}_{ad} - \ldots +
  t^{b\ldots}{}_{d\ldots}C^d{}_{ac} + \ldots.
\end{equation}
In particular, it follows that if the energy per particle $\epsilon$
is a function only of scalar invariants of $k^a{}_b$, the spatial
divergence of the pressure tensor that occurs in Euler's equations
\refeq{divp} can be reexpressed as
\begin{equation}\lbeq{divpHadamard}
  D_b p^{ab} = A^{ab}{}_{cd}C^{cd}{}_b, 
\end{equation}
where $A^{ab}{}_{cd}$ is the relativistic \emph{Hadamard elasticity
tensor}
\begin{equation}
  A^{ab}{}_{cd} = 2\frac{\partial p^{ab}}{\partial g^{cd}} - p^{ab}h_{cd} -
  h^a{}_c p^b{}_d.
\end{equation}
The first two terms on the right hand side of the above equation is
the relativistic \emph{elasticity tensor},
\begin{equation} \lbeq{elast}
  E^{ab}{}_{cd} = 2\frac{\partial p^{ab}}{\partial g^{cd}} - p^{ab}h_{cd}, 
\end{equation}
which, using eq.\ \refeq{dndh}, can be written in the more compact
form
\begin{equation} 
  E^{ab}{}_{cd} = 2n\frac{\partial}{\partial g^{cd}}
  \left(\frac{p^{ab}}{n}\right). 
\end{equation}
With all indices raised, it can also be expressed in the familiar form
(\cf CQ)
\begin{equation}
  E^{abcd} = 4n\frac{\partial^2\epsilon}{\partial g_{ab} \partial
  g_{cd}}, 
\end{equation}
which follows from eq.\ \refeq{pressuretensor} and 
\begin{equation}
  \frac{\partial}{\partial g_{ab}} = 
  -g^{ac}g^{bd}\frac{\partial}{\partial g^{cd}}.  
\end{equation}
Both $A^{abcd}$ and $E^{abcd}$ are flowline orthogonal tensors
symmetric under interchange of the index pairs $ab$ and $cd$, \ie
\begin{equation}
  A^{abcd} = A^{cdab}, \quad E^{abcd} = E^{cdab}. 
\end{equation}
In addition, the elasticity tensor has the symmetry
\begin{equation}
  E^{abcd} = E^{(ab)(cd)}. 
\end{equation}
The bottom line of this discussion is that the equations of motion are
given by the Euler equations \refeq{divp} alone, which, using eq.\
\refeq{divpHadamard}, can be expressed as
\begin{equation} \lbeq{eulereq}
  (\rho h^{ab}+p^{ab})\dot{u}_b+A^{abcd}C_{cbd} = 0.
\end{equation}
This fact will be used in section \ref{sec:sound} when we examine wave
propagation in elastic media. 

Before closing this section we would like to make a digression and
argue that the Euler equations in the form \refeq{eulereq}, which to
our knowledge has not appeared before in the literature, is valid
under more general circumstances than were assumed when deriving it
above. The assumption more restrictive than necessary was that the
energy per particle $\epsilon$ is a function only of scalar invariants
of $k^a{}_b = g^{ac}k_{cb}$ with $k_{cb}$ being the pull-back of a
fixed ($n$-independent) material space metric $k_{CB}$.  Indeed, the
derivation of eq.\ \refeq{eulereq} can be immediately generalised to
the case when there exists a torsion-free connection $\tilde{D}_A$ on
$X$ such that
\begin{equation}\lbeq{covconst}
  \tilde{D}_A t_{B\ldots} = 0,
\end{equation}
for all fixed covariant tensors $t_{A\ldots}$ on $X$ which contribute
to the equation of state. This connection can be carried over into a
projected connection $\tilde{D}_a$ on $M$ in exactly the same manner
as before, \ie through eqs.\ \refeq{Dtildet} - \refeq{Dtildeimpose}.
Hence $\tilde{D}_a$ will again be uniquely determined in terms of
$D_a$ and an elasticity difference tensor $C^c{}_{ab}$ which, due to
the requirement that $\tilde{D}_A$ be torsion-free, has the index
symmetry \refeq{Csymm}. Since eqs.\ \refeq{Dtildepushfwd} and
\refeq{Dtildepullback} will hold also in this case, the pull-back of
eq.\ \refeq{covconst} is of course
\begin{equation}
  \tilde{D}_a t_{b\ldots} = 0, \quad t_{b\ldots} = \psi^*t_{B\ldots}, 
\end{equation}
which implies that the remaining steps that lead up to the Euler
equations in the form \refeq{eulereq} are still valid.

The most physically relevant situation when this generalization
applies is when the reference structure of the material space is
completely given by three \emph{crystal axis vectors} 
\begin{equation}
  E_{\Lambda}{}^A,  \quad \Lambda = 1, 2, 3,
\end{equation}
which, mathematically, are three linearly independent vector fields
assumed to be commuting, 
\begin{equation}\lbeq{Ecommute}
  [E_\Lambda,E_\Sigma]^A = 0. 
\end{equation}
Physically, the crystal axis vectors are to be thought of as
displacement vectors pointing from one lattice point to three of its
nearest neighbouring lattice points, with the three neighbours chosen
in an ordered way such that the crystal axis vectors become smooth
vector fields when going to the continuum limit. The two-dimensional
surfaces that, due to the commutation requirement \refeq{Ecommute}, are
guaranteed to be formed by any pair of crystal axis vectors 
$(E_\Lambda{}^A,E_\Sigma{}^A)$ with $\Lambda\neq\Sigma$ are consequently to be
thought of as crystal planes. The basis of one-form fields
$\omega^\Lambda{}_A$ that are dual to the crystal lattice vectors, \ie
satisfying
\begin{equation}\lbeq{reciprocaldef}
  \omega^\Lambda{}_A E_\Sigma{}^A = \delta^\Lambda{}_\Sigma, 
\end{equation}
will here be called the \emph{reciprocal lattice one-forms}, as they
are in direct correspondence with the reciprocal lattice vectors of
classical lattice/crystal theory (\cf \cite{kittel:solid}) where, of
course, no distinction is made between vectors and one-forms. However,
it should be stressed that a factor $2\pi$ should be inserted on the
right hand side of eq.\ \refeq{reciprocaldef} to conform to the
standard definition of reciprocal lattice vectors. Now, if each
lattice point is associated with a number $N_f$ of particles belonging
to the species that one wishes to consider as fundamental (\eg
baryons), the particle density form is simply
\begin{equation}\lbeq{ncell}
  n_{ABC} = 3!\,N_f\,\omega^1{}_{[A}\,\omega^2{}_B\,\omega^3{}_{C]}. 
\end{equation}
Indeed, this three-form associates a volume $N_f$ to a cell spanned by
the crystal axis vectors,
\begin{equation}
  E_1{}^A E_2{}^B E_3{}^C n_{ABC} = N_f,
\end{equation}
and, clearly, such cells are in one-to-one correspondence with the
lattice points. 

If we assume that no further tensor fields are introduced on $X$, the
energy per particle $\epsilon$ can only depend on the scalars
\begin{equation}
  I^{\Lambda\Sigma} = g^{ab}\omega^\Lambda{}_a\omega^\Sigma{}_b, \quad
  \omega^\Lambda{}_a = \psi^*\omega^\Lambda{}_A,
\end{equation}
of which six are independent since $I^{\Lambda\Sigma} = I^{\Sigma\Lambda}$. We thus
would like to introduce a torsion-free connection $\tilde{D}_A$ for
which all three reciprocal lattice one-forms are covariantly constant,
\begin{equation}\lbeq{omegacovconst}
  \tilde{D}_A\omega^\Lambda{}_B = 0. 
\end{equation}
This is possible because eqs.\ \refeq{Ecommute} and
\refeq{reciprocaldef} imply that these one-forms are closed, thus
making the antisymmetric part of eq.\ \refeq{omegacovconst}
identically satisfied. Consequently, eq.\ \refeq{omegacovconst}
implies that $\tilde{D}_A$ is the unique flat torsion-free connection
for which the coordinate frame connection coefficients vanishes when
using coordinates $x^\Lambda$ such that $\omega^\Lambda{}_A = \partial_A
x^\Lambda$. The elasticity difference tensor appearing in the Euler
equations \refeq{eulereq} will in this case take the very simple form
\begin{equation}\lbeq{Clattice}
  C^c{}_{ab} = \sum_{\Lambda=1}^3 E_\Lambda{}^c D_a\,\omega^\Lambda{}_b,
\end{equation}
where the spacetime vector fields $E_\Lambda{}^a$ can be identified with
the material space vector fields $E_\Lambda{}^A$ since they are defined by 
\begin{equation}
  \omega^\Lambda{}_a E_\Sigma{}^a = \delta^\Lambda{}_\Sigma, \quad u_a E_\Lambda{}^a = 0. 
\end{equation}
Indeed, they of course satisfy
\begin{equation}
  \psi_* E_\Lambda{}^a = E_\Lambda{}^A. 
\end{equation}
To arrive at eq.\ \refeq{Clattice} we have used that 
\begin{equation}
  D_{[a}\omega^\Lambda{}_{b]} = 0, 
\end{equation}
which follows from the antisymmetric part of eq.\
\refeq{omegacovconst} and guarantees that the symmetry property
\refeq{Csymm} holds. The result that the matter equations of motion 
in this case reduces to eq.\ \refeq{eulereq} with $C^c{}_{ab}$ given
by eq.\
\refeq{Clattice} is independent of whether or not the unsheared 
crystal structure is conformally rescaled under variation of the
particle density, \ie whether or not the compressional distortion
tensor $\tau_{AB}$ is identically vanishing. However, if we do assume
that $\tau_{AB} = 0$, we have a preferred fixed ($n$-independent)
material space metric which by necessity is of the form
\begin{equation} 
  k_{AB} = \sum_{\Lambda,\Sigma=1}^3
  k_{\Lambda\Sigma}\,\omega^\Lambda{}_A\omega^\Sigma{}_B,
\end{equation}
where the metric components $k_{\Lambda\Sigma}$ are constants subject
to the condition
\begin{equation}
  \det{(k_{\Lambda\Sigma})} = N_f^{\,\,2},
\end{equation}
thus making $n_{ABC}$ the associated volume form of $k_{AB}$. The
reason why $k_{AB}$ can only be of this form is simply because there
is no other way to form a fixed metric on $X$ without introducing
further tensor fields contributing to the equation of state. The
constants $k_{\Lambda\Sigma}$ give the lattice structure of the material in
its unsheared state. For instance, the special case 
\begin{equation}
  k_{\Lambda\Sigma} = N_f^{\,\,2/3}\delta_{\Lambda\Sigma}
\end{equation}
corresponds to cubic lattices. The three independent scalar invariants
that can be formed from $k^a{}_b$ will obviously be functions of the
six independent scalar invariants $I^{\Lambda\Sigma}$. If $\epsilon$ does not
depend on any combinations of the latter that cannot be formed out of
$k^a{}_b$, the equation of state belongs to the class that this paper
is mainly concerned with. Incidentally, this serves as a good
motivation for taking the material space metric $k_{AB}$ to be flat.

\section{Eigenvalue and eigenvector formulation}
\label{sec:eig}

In this section we shall explore the fact that whichever scalar
invariants of $k^a{}_b$ we are using to parametrise the equation of
state by, they can always be expressed in terms of the eigenvalues of
$k^a{}_b$ and, consequently, so can the energy per particle
$\epsilon$. Since $k^a{}_b$ is flowline orthogonal and since $h_{ab}$
(viewed as a three-dimensional metric) is positive definite, the
eigenvectors of $k^a{}_b$ corresponding to distinct eigenvalues are
automatically orthogonal, while eigenvectors with the same eigenvalue
can be chosen orthogonal.  Moreover, since $k_{ab}$ is also positive
definite, all eigenvalues of $k^a{}_b$ will be positive and we shall
write them as $n_\mu^{\,\,2}$, $\mu = 1,2,3$, in terms of their positive
square roots $n_\mu$ which, in a loose sense, represent
one-dimensional particle densities in the principal directions, \ie
along the eigenvectors. We hence refer to these quantities as
\emph{principal linear particle densities}, or \emph{linear particle
  densities} for short.  According to eq.\ \refeq{ndet}, the particle
density $n$ is simply the product of the linear particle densities,
\ie 
\begin{equation}
  n = n_1 n_2 n_3.
\end{equation}
Using an orthonormal basis $\{e_\mu{}^a \}$ of eigenvectors of
$k^a{}_b$, implying that
\begin{align}
  &g_{ab} = -u_au_b + \sum_{\mu=1}^3 e_{\mu\,a}e_{\mu\,b}, \\ 
  &k_{ab} = \sum_{\mu=1}^3 n_\mu^{\,\,2}\,e_{\mu\,a}e_{\mu\,b},
\end{align}
it follows from eq. \refeq{dellhab} that the operator
$\partial/\partial g^{ab}$, when acting on scalar invariants of
$k^a{}_b$, takes the particularly simple form
\begin{equation}\lbeq{ddheigen}
 \frac{\partial}{\partial g^{ab}} =
  \tsfrac12\sum_{\mu=1}^3
  e_{\mu\,a}e_{\mu\,b}\,n_{\mu}\frac{\partial}{\partial n_{\mu}}. 
\end{equation}
The pressure tensor then becomes
\begin{equation}
  p_{ab} = \sum_{\mu=1}^3 p_\mu\,e_{\mu\,a}e_{\mu\,b},
\end{equation}
where
\begin{equation}\lbeq{peigen}
  p_\mu = n\,n_\mu\frac{\partial\epsilon}{\partial n_\mu}. 
\end{equation}
Hence $p^a{}_b$ has the same eigenvectors as $k^a{}_b$, while the
eigenvalues $p_\mu$, which we call the \emph{principal pressures},
are given by eq.\ \refeq{peigen}.

We now proceed to calculate the elasticity tensor according to eq.\
\refeq{elast}. To do so we need to know how to differentiate the
eigenvectors with respect to $g^{ab}$. Therefore we let
$\partial/\partial g^{ab}$ act on the relations
\begin{align}
 g_{cd}\,e_\rho{}^ce_\sigma{}^d &= \delta_{\rho \sigma}, \\
 k_{cd}\,e_\rho{}^ce_\sigma{}^d &= n_\rho^{\,\,2}\,\delta_{\rho \sigma}.
\end{align}
This results in, respectively,
\begin{align}
 e_{\rho\,c}\frac{\partial e_\sigma{}^c}{\partial g^{ab}}+
 e_{\sigma\,c}\frac{\partial e_\rho{}^c}{\partial g^{ab}} 
 &= e_{\rho (a}e_{\sigma\,b)} \\
 n_\rho^{\,\,2}\,e_{\rho\,c}\frac{\partial e_\sigma{}^c}{\partial g^{ab}} + 
 n_\sigma^{\,\,2}\,e_{\sigma\,c}\frac{\partial e_\rho{}^c}{\partial g^{ab}}
  &= n_\rho^{\,\,2}\,e_{\rho\,a}e_{\rho\,b}\,\delta_{\rho \sigma} 
\end{align}
Assuming that $n_\rho\neq n_\sigma$ when $\rho\neq\sigma$, we can
solve for all projections of the differentiated basis vectors. The
result reads
\begin{align}
  e_{\rho\,c}\frac{\partial e_\rho{}^c}{\partial g^{ab}} &=
  \tsfrac12\,e_{\rho (a}e_{\rho\,b)} \\ 
  e_{\sigma\,c}\frac{\partial e_\rho{}^c}{\partial g^{ab}} &=
  \frac{n_\rho^{\,\,2}}{n_\rho^{\,\,2}-n_\sigma^{\,\,2}}\,e_{\rho
  (a}e_{\sigma\,b)} \quad \rho\neq\sigma,
\end{align}
which implies
\begin{equation}\lbeq{dedh}
  \frac{\partial e_\rho{}^c}{\partial g^{ab}} = \tsfrac12\,
  e_\rho{}^ce_{\rho\,a}e_{\rho\,b} +
  \sum_{\sigma\neq\rho}\frac{n_\rho^{\,\,2}}{n_\rho^{\,\,2}-n_\sigma^{\,\,2}}\,
  e_\sigma{}^c e_{\rho(a}e_{\sigma\,b)}.
\end{equation}
Making use of eq.\ \refeq{ddheigen} when $\partial/\partial g^{ab}$
acts on the scalars $p_\rho$, we find
\begin{equation}
  E_{ab}{}^{cd} = \sum_{\rho=1}^3 \left[ 2p_\rho
    \left(e_\rho{}^c\frac{\partial e_\rho{}^d}{\partial g^{ab}} +
      e_\rho{}^d\frac{\partial e_\rho{}^c}{\partial g^{ab}} \right) +
    \sum_{\sigma=1}^3\left(n_\sigma\frac{\partial p_\rho}{\partial
        n_\sigma}
      - p_\rho\right) e_{\sigma a}e_{\sigma
    b}e_\rho{}^ce_\rho{}^d\right].
\end{equation}
Using eq.\ \refeq{dedh} for the derivatives of the basis vectors and
raising the indices $a$ and $b$, we obtain our final form for the
elasticity tensor:
\begin{equation}
  E^{abcd} = \sum_{\rho=1}^3 \left[2p_\rho\, e_\rho{}^a e_\rho{}^b
  e_\rho{}^c e_\rho{}^d + 2\sum_{\sigma\neq\rho}
  \frac{n_\rho^{\,2}p_\rho-n_\sigma^{\,2}p_\sigma}{n_\rho^{\,2}-
  n_\sigma^{\,2}}\,e_\rho{}^{(a}e_\sigma{}^{b)}e_\rho{}^{(c}e_\sigma{}^{d)}
  + \sum_{\sigma=1}^3\left(n_\sigma\frac{\partial p_\rho}{\partial
  n_\sigma}-p_\rho\right)e_\sigma{}^ae_\sigma{}^b e_\rho{}^c
  e_\rho{}^d \right].
\end{equation}
Note that the symmetry property $E^{abcd} = E^{cdab}$ requires that
the quantities
\begin{equation}
  n_\sigma\frac{\partial p_\rho}{\partial n_\sigma}-p_\rho
\end{equation}
are symmetric under $\rho\leftrightarrow\sigma$. This is indeed the
case, since, according to eq.\ \refeq{peigen}, they can be expressed as
\begin{equation}
  n_\sigma\frac{\partial p_\rho}{\partial n_\sigma}-p_\rho = 
  n\,n_\rho n_\sigma\frac{\partial^2\epsilon}{\partial n_\rho \partial
  n_\sigma} + n\,n_\rho\frac{\partial\epsilon}{\partial n_\rho}\,\delta_{\rho\sigma}. 
\end{equation}
The Hadamard elasticity tensor is now
\begin{equation}
  A^{abcd} = \sum_{\rho=1}^3\left\{ \sum_{\sigma=1}^3
  n_\sigma\frac{\partial p_\rho}{\partial n_\sigma}\,e_\sigma{}^a
  e_\sigma{}^b e_\rho{}^c e_\rho{}^d + 2\sum_{\sigma\neq\rho} \left[
  \frac{n_\rho^{\,2}p_\rho-n_\sigma^{\,2}p_\sigma}{n_\rho^{\,2}-
  n_\sigma^{\,2}}\,e_\rho{}^{(a}e_\sigma{}^{b)}e_\rho{}^{(c}e_\sigma{}^{d)}
  - p_\rho\,e_\sigma{}^a e_\rho{}^d e_\rho{}^{(b}e_\sigma{}^{c)}
  \right] \right\}.
\end{equation}

Let us now return to the decomposition $k^a{}_b = n^{2/3}\eta^a{}_b$,
which, as discussed in section \ref{sec:relasticity}, in a natural way splits
$p_{ab}$ into its trace part $p\lagom h_{ab}$ and its trace-free part
$\pi_{ab}$. Doing so, we do no longer wish to parametrise the equation
of state by the linear particle densities $n_\mu$, but rather by the
particle density $n$ and variables that are closely related to the
eigenvalues of $\eta^a{}_b$. Clearly, $\eta^a{}_b$ has the same
orthonormal eigenvectors $e_\mu{}^a$ as $k^a{}_b$, the eigenvalues
being $(\alpha_1^{\,2},\alpha_2^{\,2},\alpha_3^{\,2})$, say, with
\begin{equation}
  \alpha_\mu = \frac{n_\mu}{n^{1/3}} =
  \left(\frac{n_\mu^2}{n_{\mu+1}n_{\mu+2}}\right)^{\!\!1/3} =
  \left(\frac{z_{\mu+2}}{z_{\mu+1}}\right)^{\!\!1/3}, \quad
  \alpha_1\alpha_2\alpha_3 = 1.
\end{equation}
where
\begin{equation}
  z_\mu = \frac{n_{\mu+1}}{n_{\mu+2}} =
  \frac{\alpha_{\mu+1}}{\alpha_{\mu+2}}, \quad z_1z_2z_3 = 1,
\end{equation}
and the cyclic rule $\mu+3=\mu$ is being used. We find it useful to
parametrise scalar invariants of $\eta^a{}_b$ by the $z_\mu\!$'s. From
$n=n_1n_2n_3$ and $z_\mu = n_{\mu+1}/n_{\mu+2}$ it follows that
\begin{equation}\lbeq{ddnddz}
  n_\mu\frac{\partial}{\partial n_\mu} = n\frac{\partial}{\partial n}
  + z_{\mu+2}\frac{\partial}{\partial z_{\mu+2}} -
  z_{\mu+1}\frac{\partial}{\partial z_{\mu+1}},
\end{equation}
which via eqs.\ \refeq{ddhneta} and \refeq{ddheigen} gives
\begin{equation}
  \eta_{c<a}\frac{\partial}{\partial\eta^{b>}{}_c} =
  \tsfrac12\sum_{\mu=1}^3
  e_{\mu\,a}e_{\mu\,b}\left(z_{\mu+2}\frac{\partial}{\partial
  z_{\mu+2}} - z_{\mu+1}\frac{\partial}{\partial z_{\mu+1}} \right).
\end{equation}
It hence follows from eq.\ \refeq{pieta} that the anisotropy pressure
tensor is
\begin{align}\lbeq{anisp}
  \pi_{ab} &= \sum_{\mu=1}^3\pi_\mu e_{\mu\,a}e_{\mu\,b}, \\
    \pi_\mu &= n\left( z_{\mu+2}\,\frac{\partial\epsilon}{\partial
  z_{\mu+2}} - z_{\mu+1}\,\frac{\partial\epsilon}{\partial z_{\mu+1}}
  \right),
\end{align}
where the eigenvalues $\pi_\mu$ obey $\sum_{\mu=1}^3\pi_\mu=0$
and are related to the principal pressures $p_\mu$ and the isotropic
pressure $p$ by
\begin{equation}
  p_\mu = p + \pi_\mu.
\end{equation}
Note that the combinations $z_{\mu+2}\,\partial/\partial z_{\mu+2} -
z_{\mu+1}\,\partial/\partial z_{\mu+1}$ all annihilate $z_1z_2z_3$,
which means that the condition $z_1z_2z_3=1$ may be used to simplify
anything these operators act on.

\section{Wave propagation in elastic materials}\label{sec:sound}%

Our discussion of the speed of sound in elastic materials will be
based on the discontinuity formalism introduced by Hadamard about a
century ago. The notation will follow Carter's as presented in
\cite{carter:sound}, but as our starting point we will take the
Euler equations expressed in terms of the Hadamard tensor, see eq.\
\refeq{eulereq}.   

\begin{figure}[ht]
 \centering \begin{minipage}[t]{0.78\linewidth} \centering
 \includegraphics[width=0.7\textwidth, angle=0]{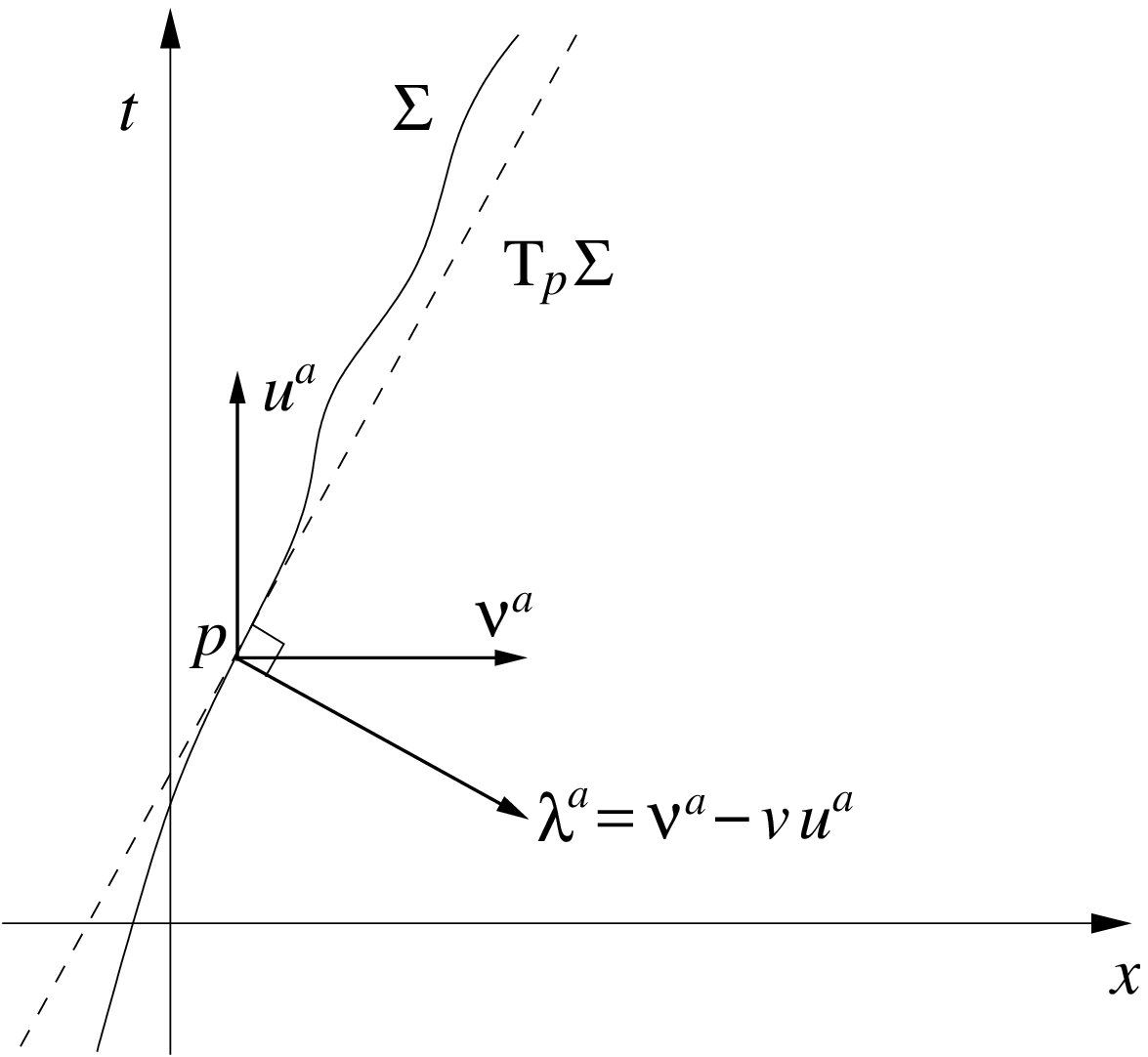}
 \caption{This schematic picture shows the wave front $\Sigma$ moving
 in spatial direction $\nu^a$ with speed $v$ with respect to the flow
 $u^a$ at a point $p$. It is convenient to introduce the vector
 $\lambda^a = \nu^a-vu^a$ which is normal to $\Sigma$ and gives the
 wave propagation speed as its contraction with $u^a$. Also sketched
 by a dashed line is the tangent space of $\Sigma$ at $p$.}
\label{fig:wavefront} \end{minipage}
\vspace{120pt}
\end{figure}

We consider a hypersurface $\Sigma$ representing a
wave front across which second order derivatives of the material space
mapping $\psi$ are allowed to be discontinuous, whereas continuity is
assumed for all first order derivatives of $\psi$ as well as of the
spacetime metric. In other words the assumption is that $\psi$ and
$g_{ab}$ are both $C^1$. The metric may be assumed to have a higher
degree of differentiability, but this is irrelevant to the
discussion. The continuity of the first order derivatives of $\psi$
implies that all tensors on $M$ which are pulled back from $X$  are
continuous.  Since it thus follows that the particle four-velocity
$u^a$ is continuous,
\ie well-defined on $\Sigma$, the wave front propagates at a
well-defined speed $v$ with respect to the particle flow, see fig.\ 
\ref{fig:wavefront}.

In order to derive an equation governing the propagation of the wave
front we start by noting that the tensors $\rho h^{ab}+p^{ab}$ and
$A^{abcd}$ occurring in Euler's equations are continuous, whereas the
acceleration $\dot{u}^a$ and the elasticity difference tensor
$C^c{}_{ab}$ may be discontinuous across $\Sigma$. Denoting the
discontinuity of a quantity by surrounding square brackets we
therefore have
\begin{equation}\lbeq{discoeuler}
    (\rho h^{ab}+p^{ab})[\dot{u}_b]+A^{abcd}[C_{cbd}] = 0.
\end{equation}
Following Carter, we set 
\begin{equation} \lbeq{discudot}
  [\dot{u}_a]=\alpha \iota_a, \quad \iota^a\iota_a=1,
\end{equation}
where $\alpha$ is the \emph{amplitude}  and $\iota_a$ is the
\emph{polarization vector} of the wave front.  Since the discontinuity
of the gradient of a scalar is always normal to the wave front and the
Christoffel symbols are continuous due to the continuity of the first
order metric derivatives it follows that for any tensor
$t^{a\ldots}{}_{b\ldots}$ defined across $\Sigma$, there exists a
tensor $\tau^{a\ldots}{}_{b\ldots}$ defined on $\Sigma$ such that
\begin{equation}
  [\nabla_c t^{a\ldots}{}_{b\ldots}] =
  \tau^{a\ldots}{}_{b\ldots}\lambda_c, \quad \lambda_a=\nu_a-vu_a,
\end{equation}
where $\nu_a$ is the \emph{propagation vector} of the wave front,
assumed to be normalised,
\begin{equation}
  \nu^a\nu_a = 1. 
\end{equation}
 It follows that there are tensors $\kappa^a$ and $\zeta_{ab}$ such
 that
\begin{align}
  [\nabla_b u^a] &= \kappa^a\lambda_b \lbeq{discdu} \\
  [\nabla_c k_{ab}] &= \zeta_{ab}\lambda_c. \lbeq{discdk}
\end{align}
By contracting \refeq{discdu} with $u^b$ and using eq.\ \refeq{discudot} we find
\begin{equation}
   \kappa^a=\frac{\alpha}{v}\iota^a 
\end{equation}
Moreover, the discontinuity of $\mathcal{L}_u k_{ab} = u^c\nabla_ck_{ab} + 2 k_{c(a}\nabla_{b)}u^c = 0$ gives
\begin{equation}
  \zeta_{ab} = -2\frac{\alpha}{v^2}\,\iota^c k_{c(a}\lambda_{b)},
\end{equation}
which in turn implies
\begin{equation}\lbeq{discC}
  [C^c{}_{ab}] = -\frac{\alpha}{v^2}\,\iota^c\nu_a\nu_b.
\end{equation}
Finally, substituting eqs.\ \refeq{discudot} and \refeq{discC} into
eq.\ \refeq{discoeuler} we arrive at
\begin{equation}\lbeq{chareq}
    [v^2(\rho h^{ac}+p^{ac})-Q^{ac}]\iota_c = 0,
\end{equation}
where $Q^{ac} = Q^{(ac)}$ is the relativistic \emph{Fresnel tensor},
given by
\begin{equation}
    Q^{ac} = A^{abcd}\nu_b\nu_d.
\end{equation}
We have now shown that the Euler equations constrain the quantities
$\nu_a$, $\iota_a$ and $v$ algebraically according to eq.\
\refeq{chareq}. In realistic situations the combination $\rho h^{ab}+p^{ab}$, 
viewed as a three-dimensional tensor, is expected to be
non-degenerate. Clearly, the propagation vector $\nu_a$ can then be
specified freely since for each $\nu_a$ eq.\
\refeq{chareq} turns into a three-dimensional eigenvalue problem, with
$v^2$ being the eigenvalue and $\iota_a$ the eigenvector. Since eq.\
\refeq{chareq} is a three-dimensional equation, it has three
independent solutions. Moreover it may be noted that as a consequence
of the symmetry of $Q^{ab}$, it follows that any two polarization
vectors ${}^{(1)}\iota_a$ and ${}^{(2)}\iota_a$ corresponding to
distinct eigenvalues ${}^{(1)}v^2$ and ${}^{(2)}v^2$ are orthogonal with
respect to $\rho h^{ab} + p^{ab}$. For a general propagation vector
$\nu^a$, the Fresnel tensor can be written as
\begin{equation}
\begin{split}
  Q^{ac} &= \sum_{\rho=1}^3 \left\{ \sum_{\sigma=1}^3
  \nu_\rho\nu_\sigma\,n_\sigma\frac{\partial p_\rho}{\partial n_\sigma}\,
  e_\rho{}^{(a}e_\sigma{}^{c)} + \sum_{\sigma\neq\rho}
  \frac{n_\sigma^{\,\,2}(p_\sigma-p_\rho)}{n_\sigma^{\,\,2}-n_\rho^{\,\,2}}
  \left[ \nu_\rho^{\,\,2}e_\sigma{}^a e_\sigma{}^c + \nu_\rho \nu_\sigma
  e_\rho{}^{(a}e_\sigma{}^{c)} \right] \right\} \\
  &= \sum_{\rho=1}^3 \left[ \nu_\rho^{\,\,2}n_\rho\frac{\partial
  p_\rho}{\partial n_\rho} + \sum_{\sigma\neq\rho} \nu_\sigma^{\,\,2}
  \frac{n_\rho^{\,\,2}(p_\rho-p_\sigma)}{n_\rho^{\,\,2}-n_\sigma^{\,\,2}}
  \right] e_\rho{}^a e_\rho{}^c \\
  &+ \sum_{\rho=1}^3\sum_{\sigma\neq\rho} v_\rho v_\sigma \left[
  n_\sigma\frac{\partial p_\rho}{\partial n_\sigma} +
  \frac{n_\sigma^{\,\,2}(p_\sigma-p_\rho)}{n_\sigma^{\,\,2}-n_\rho^{\,\,2}}
  \right] e_\rho{}^{(a}e_\sigma{}^{c)},
\end{split}
\end{equation}
where $\nu_\rho = e_\rho{}^a\nu_a$. We shall only solve the
characteristic equation explicitly for the case when the propagation
vector $\nu_a$ coincides with one of the eigenvectors, say
$e_{\mu a}$. The Fresnel tensor then reduces to
\begin{equation}
  Q^{ac} = n_\mu\frac{\partial p_\mu}{\partial n_\mu}\,e_\mu{}^a
  e_\mu{}^c +
  \sum_{\sigma\neq\mu}\frac{n_\sigma^{\,2}(p_\sigma-p_\mu)}{n_\sigma^{\,2}-n_\mu^{\,2}}\,e_\sigma{}^a e_\sigma{}^c.
\end{equation}
Since
\begin{equation}
  \rho h^{ac} + p^{ac} =
  \sum_{\sigma=1}^3(\rho+p_\sigma)e_\sigma{}^a e_\sigma{}^c,
\end{equation}
it follows that the characteristic equation \refeq{chareq} has one
solution corresponding to a purely longitudinally polarised mode,
given by
\begin{equation}\lbeq{v2long}
  v^2 = v_{\mu||}^{\,2} = (\rho+p_\mu)^{-1}\,n_\mu\ds\frac{\partial
  p_\mu}{\partial n_\mu}, \quad \iota^a = \nu^a = e_\mu{}^a.
\end{equation}
The remaining two independent solutions are 
\begin{equation}
  v^2 = v_{\mu\perp\nu}^{\,2} =
  (\rho+p_\nu)^{-1}\,\frac{n_\nu^{\,2}(p_\nu-p_\mu)}{n_\nu^{\,2}-n_\mu^{\,2}},
  \quad \iota^a = e_\nu{}^a, \quad \iota^a\nu_a=0.
\end{equation}
These modes correspond to transversal waves whose polarization vector
coincides with one of two eigenvectors $e_\nu{}^a$ orthogonal to the
propagation eigenvector $e_\mu{}^a$.  Remarkably, the formula
\refeq{v2long} for longitudinal waves in eigenvector directions
exactly mimics the corresponding formula for the speed of sound $c_s$
in a perfect fluid with equation of state $\rho=\rho(n)$, $p=p(n)$,
namely
\begin{equation}
  c_s^{\,\,2} = (\rho+p)^{-1}\,n\frac{dp}{dn}.
\end{equation}
For perfect fluids the quantity $\beta = n\,dp/dn$ can be identified
with the compressibility modulus, or bulk modulus. Thus, since a
perfect fluid can be viewed as a special case of an elastic material,
we take the analogy further by introducing the
\emph{principal compressibility moduli} $\beta_\mu$ according to
\begin{equation}
  \beta_\mu = n_\mu\frac{\partial p_\mu}{\partial n_\mu},
\end{equation}
in terms of which the speeds of the eigenvector directed longitudinal
waves (sound waves) are given by
\begin{equation}
  v_{\mu||}^{\,\,2} = \frac{\beta_\mu}{\rho+p_\mu}. 
\end{equation}
It is worth noting that $v_{\mu||}^{\,\,2}$ can also be rewritten as 
\begin{equation}
  v_{\mu||}^{\,\,2} 
  = \frac{\partial p_\mu}{\partial n_\mu}\mbox{\Large /}\frac{\partial\rho}{\partial n_\mu},
\end{equation}
analogous to the well-known perfect fluid formula $c_s^{\,\,2} = dp/d\rho$. 

Our resulting expressions for the nine principal speeds
$v_{\mu||}$ and $v_{\mu\perp\nu}$ are relevant to compare with those
obtained by Maugin in \cite{maugin:prestressed}, since in both cases
it is assumed that the equation of state is set up using a fixed
metric as the only material space structure. We could go through some
lengths to establish the precise connection between the two sets of
formulae, but refrain from doing so since it would only lead to the
introduction and explanation of new notation rather than to something
illuminating for anyone but a very specialised reader. In
\cite{maugin:prestressed} the propagation speeds are given in terms of
quantities that explicitly refers to a particular way of expressing
the pressure tensor $p_{ab}$ by making use of the Cayley-Hamilton
theorem. In contrast, our recipe for calculating $v_{\mu||}$ and
$v_{\mu\perp\nu}$ from the equation of state $\epsilon =
\epsilon(n_1,n_2,n_3)$ is very simple and direct which should make it
favourable for practical applications.

Although we do not treat the case of a general propagation direction,
it is worth noting that the principal speeds $v_{\mu||}$ and
$v_{\mu\perp\nu}$ are stationary under first order variations of the
propagation vector $\nu^a$ around the eigenvector $e_\mu{}^a$. This is
easily proved by taking the first order variation of the
characteristic equation \refeq{chareq} induced by a first order
variation $\delta\nu^a$ in $\nu^a$,
\begin{equation}\lbeq{deltachar}
  [\delta(v^2)(\rho h_{ac}+p_{ac}) -
  2A_{abcd}\,\nu^{(b}\delta\nu^{d)}]\iota^c + [v^2(\rho h_{ac} + p_{ac})
  - Q_{ac}]\,\delta\iota^c = 0.
\end{equation}
This equation is supplemented by the restrictions
\begin{equation}\lbeq{restrictiota}
  \nu_a\delta\nu^a = 0, \quad \iota_a\delta\iota^a = 0,
\end{equation}
that follows from $\nu^a$ and $\iota^a$ being unit vectors. Requiring
that $\delta(v^2) = 0$ gives four linear equations for the three
components of $\delta\iota^a$, namely eqs.\ \refeq{deltachar} and
\refeq{restrictiota}. 
In order for the system not to be overdetermined these equations 
cannot be linearly independent.
To prove that we do
have linear dependence in the case when $\nu^a$ is an eigenvector of
$k^a{}_b$ we contract eq.\ \refeq{deltachar} with $\iota^a$ which
gives the condition
\begin{equation}\lbeq{Bnudnu}
  B_{bd}\nu^b\delta\nu^d = 0,
\end{equation}
where $B_{bd}$ is the symmetric tensor
\begin{equation}
  B_{bd} = A_{abcd}\iota^a\iota^c = E_{abcd}\iota^a\iota^c
  - p_{bd}.  
\end{equation}
As we have seen, taking $\nu^a$ to be an eigenvector of $k^a{}_b$
leads to $\iota^a$  being an eigenvector as well. In this case
$B_{bd}$ is diagonal in the eigenvector basis, leading to
\refeq{Bnudnu} being automatically satisfied. Thus, the number 
of independent equations for $\delta\nu^a$ is at most three, which
proves our statement that $v$ is stationary under variations of
propagation direction around a principal direction.


\section{A particular equation of state}\label{sec:eos}

In this work will assume an equation of state where the energy per
particle depends on only one invariant of $\eta^a{}_b$, namely the
\emph{shear scalar}
\begin{equation}
s^2 =
\tsfrac1{36}\left[ (\eta^a{}_a)^3 -
\eta^a{}_b\eta^b{}_c\eta^c{}_a-24 \right] =
\tsfrac12\sum_{\mu=1}^3s_\mu^{\,\,2}, 
\quad s_\mu^{\,\,2} = \tsfrac16(z_\mu^{-1}-z_\mu)^2,  
\end{equation}
where, as before, $z_\mu = n_{\mu+2}/n_{\mu+1}$. This definition of
$s^2$ differs from that of Carter and Quintana, for reasons that are
discussed in the appendix. As also discussed there, the equation of
state is assumed to be of the form
\begin{equation}
\epsilon = \check{\epsilon}+\frac{\check{\mu}}{n}s^2,
\end{equation}
where checked symbols denotes that the quantity is unsheared, \ie 
depends on $n$ only. The quantity $\check\epsilon$ is the unsheared
energy per particle that was discussed in section
\ref{sec:relasticity}. The quantity $\check\mu$, on the other hand, is
the \emph{shear modulus} or \emph{modulus of rigidity}. Equivalently,
the equation of state can be given as
\begin{equation}
  \rho = \check\rho + \sigma,
\end{equation}
with $\check\rho = n\check\epsilon$ being the \emph{unsheared energy
  density} and $\sigma = \check\mu s^2$ being the \emph{shearing
  energy density}.

For this class of equations of states, the pressure tensor $p_{ab}$
takes the form
\begin{align}
  &p_{ab} = p\,h_{ab} + \pi_{ab},
\end{align}
where
\begin{align}
  &p = \check{p} +
  (\check\Omega-1)\sigma, \quad \check{p} =
  n^2\frac{d\check{\epsilon}}{dn}, \quad \check{\Omega} =
  \frac{n}{\check{\mu}}\frac{d\check{\mu}}{dn} \\ \lbeq{phateq}
  &\pi_{ab} = \tsfrac16 \check{\mu}\left[(\eta^c{}_c)^2\eta_{<ab>} -
  \eta^{cd}\eta_{c<a}\eta_{\,b>d}\right].
\end{align}
The quantity $\check{p}$ will be referred to as the \emph{unsheared
pressure}. In the eigenvector basis the anisotropy pressure tensor
$\pi_{ab}$ is given by eq.\ \refeq{anisp}, the eigenvalues being
\begin{equation}
  \pi_\mu = \check\mu\left(\chi_{\mu+1}-\chi_{\mu+2}\right), \quad
  \chi_\nu = \tsfrac16(z_\nu^{-2}-z_\nu^{\,2}).
\end{equation}
It is worth noting that the $z_\mu$-dependent quantities
$s_\mu^{\,\,2}$ and $\chi_\mu$ satisfy
\begin{equation}
  z_\mu\frac{\partial s_\mu^{\,2}}{\partial z_\mu} = -2\chi_\mu, \quad
  z_\mu\frac{\partial\chi_\mu}{\partial z_\mu} = -2(s_\mu^{\,2}+\tsfrac13). 
\end{equation}
This facilitates the calculation of the principal compressibility
moduli, which we find are given by
\begin{equation}
  \beta_\mu = \beta + 4\sigma - 2\sigma_\mu + (2\check\Omega-1)\pi_\mu, 
\end{equation}
where we have introduced the shorthand notation
\begin{equation}
\beta = n\frac{\partial p}{\partial n} = \check\beta + \tsfrac43\check\mu 
 + \left[ (\check\Omega-1)\check\Omega +
 \check\beta\,\frac{d\check\Omega}{d\check{p}} \right] \sigma.
\end{equation}
Here, the quantity $\check\beta$ is the \emph{bulk modulus}
\begin{equation}
  \check\beta = n\frac{d\check{p}}{dn} =
  (\check\rho+\check{p})\frac{d\check{p}}{d\check\rho},
\end{equation}
and we have used $n\,d\check\Omega/dn = \check\beta
\,d\check\Omega/d\check{p}$. Our identification of $\check\beta$ and
$\check\mu$ as the bulk and shear moduli is justified in the
appendix. As seen in section \ref{sec:sound}, the principal
longitudinal speeds of sound are now given by
\begin{equation}
  v_{\mu||}^{\,\,2} = \frac{\beta_\mu}{\rho+p_\mu}.
\end{equation}
On the other hand, the speeds of the principal transversal waves are
straightforwardly calculated to be
\begin{equation}
  v_{\mu\perp\nu}^{\,\,2} = \frac{\check\mu + 2\sigma -
  \tsfrac12(\sigma_\mu + \sigma_\nu + \pi_\mu-\pi_\nu)}{\rho+p_\nu}.
\end{equation}
Note that in the limit of zero shear, $z_\mu \rarr 1$, the principal
speeds goes over into
\begin{align}
  &v_{\mu||}^{\,\,2} \rarr v_{||}^{\,\,2} = \frac{\check\beta +
  \tsfrac43\check\mu}{\check\rho+\check{p}} =
  \frac{d\check{p}}{d\check\rho} +
  \frac43\frac{\check\mu}{\check\rho+\check{p}} \\
  &v_{\mu\perp\nu}^{\,\,2} \rarr v_\perp^{\,\,2} =
  \frac{\check\mu}{\check\rho+\check{p}},
\end{align}
in agreement with Carter's results \cite{carter:sound}. 

\section{The degenerate case of two equal eigenvalues and orthogonal
  eigenvectors}\label{sec:deg}

In this section we shall summarise some useful relations for the
degenerate case when two of the eigenvalues of $k^a{}_b$ are equal,
\ie when two of the linear particle densities, say $n_2$ and $n_3$, are equal. 
This is the situation one for instance encounters in the spherically
symmetric case. With this application in mind we will label quantities
corresponding to the nondegenerate and degenerate eigenvalues by $r$
(for radial) and $t$ (for tangential) respectively. We thus set
\begin{align}
  &n_r = n_1, \quad n_t = n_2 = n_3 \\
  &r_a = e_{1\,a}, \quad t_{ab} = h_{ab} - r_a r_b = e_{2\,a}e_{2\,b}
  + e_{3\,a}e_{3\,b},  
\end{align}
with $t^a{}_b$ being the projection tensor that projects onto the
degenerate eigenvector two-plane. 
The pulled back material space
metric $k_{ab}$ can now be written as
\begin{equation}\lbeq{kdeg}
  k_{ab} = n_r^{\,\,2}r_ar_b +n_t^{\,\,2}t_{ab}.
\end{equation}
When calculating the pressure tensor we may take the view that we are
dealing with an effective two-parameter equation of state
$\epsilon_{\mathrm{eff}}(n_r, n_t) = \epsilon(n_r, n_t, n_t)$. We are allowed to do so
because the full three-parameter equation of state $\epsilon(n_1, n_2,
n_3)$ is invariant under permutations of the arguments, which in turn
is due to the corresponding symmetry of the scalar invariants of
$k^a{}_b$.  Note however that, in general, we cannot use the effective
equation of state when calculating higher order derivatives of
$\epsilon$, since the first order derivatives do not obey the
permutation symmetry.  In particular, one needs the full
three-parameter equation of state to calculate the elasticity tensor,
which should not be surprising since it is used to determine the speed
of wave propagation. Using the effective equation of state the
pressure tensor may be written as
\begin{equation}
  p_{ab} = p_r\,r_a r_b + p_t\,t_{ab}, 
\end{equation}
where
\begin{align}
  p_r &= n\,n_r\frac{\partial\epsilon_{\mathrm{eff}}}{\partial n_r}, \\
  p_t &= \tsfrac12 n\,n_t\frac{\partial\epsilon_{\mathrm{eff}}}{\partial n_t}.
\end{align}

For practical purposes it is often convenient to change the two
effective equation of state parameters from $(n_r, n_t)$ to $(n, z)$,
with
\begin{equation}
   n = n_r n_t{}^{\!\!2}, \quad z = n_r/n_t. 
\end{equation}
The function $z$ represents the only independent quotient $z_\mu
=n_{\mu+1}/n_{\mu+2}$, since clearly $z_1 = 1$ and \mbox{$z_2^{\,-1} =
z_3 = z$}. Making this change of parameters we find
\begin{equation}
  k_{ab} = n^{2/3}\eta_{ab}, \quad \eta_{ab} =
  z^{1/3}(z\,r_a r_b + z^{-1}j_{ab}).
\end{equation}
Defining a new traceless tensor
\begin{equation}
  l_{ab} = -2r_a r_b + j_{ab}, 
\end{equation}
we may calculate $\pi_{ab}$ according to the very simple formula
\begin{equation}
  \pi_{ab} = q\,l_{ab},
\end{equation}
where 
\begin{equation}
  q = \pi_t = -\tsfrac12\pi_r = -\tsfrac12 nz\frac{\partial \epsilon_{\textrm{eff}}}{\partial z}
\end{equation}
Thus, by specifying $\epsilon = \epsilon_{\textrm{eff}}(n,z)$, the pressure
tensor is given by the relation
\begin{equation}
  p_{ab} = p\,h_{ab} + q\,l_{ab}, 
\end{equation}
where $p$, as before, is the isotropic pressure 
\begin{equation}
  p = n^2\frac{\partial \epsilon_{\textrm{eff}}}{\partial n}
\end{equation}
 
As pointed out above, we cannot calculate all speeds of wave
propagation using only the effective equation of state. However, for
special cases of the directions of propagation and polarization, it is
possible.  Out of the nine speeds $v_{\mu||}$ and $v_{\mu\perp\nu}$ in
the principal directions only five are independent, namely
\begin{equation}
\begin{split}
  v_{r||} &= v_{1||} \\
  v_{t||} &= v_{2||} =v_{3||} \\
  v_{r\perp} &=  v_{1\perp 2} =  v_{1\perp 3} \\
  v_{t\perp r} &=  v_{2\perp 1} =  v_{3\perp 1} \\
  v_{t\perp t} &=  v_{2\perp 3} =  v_{3\perp 2} \\
\end{split}
\end{equation}
Of these five only two actually requires the use of the full
three-parameter equation of state, namely those for which both the
propagation and polarization vectors lie in the degenerate two-plane,
\ie $v_{t||}$ and $ v_{t\perp t}$. The remaining three are given by 
\begin{align}
  &v_{r||}^{\,\,2} = \frac{\beta_r}{\rho+p_r}, \quad \beta_r =
  n_r\frac{\partial p_r}{\partial n_r} \\ 
  &v_{r\perp}^{\,\,2} =
  \frac{n_t^{\,\,2}(p_t-p_r)}{n_t^{\,\,2}-n_r^{\,\,2}} =
  \frac{3q}{1-z^2} \\ 
  &v_{t\perp r}^{\,\,2} =
  \frac{n_r^{\,\,2}(p_r-p_t)}{n_r^{\,\,2}-n_t^{\,\,2}} =
  \frac{3q}{z^{-2}-1}.   
\end{align}

Having the full three-parameter equation of state of course makes it
unnecessary to be cautious about setting $n_2=n_3$ before calculating
various quantities, since we may simply specialise the general
formulae to the degenerate case, \ie set (or take the limit) $n_2=n_3$
afterwards. Doing this for the equation of state treated in section
\ref{sec:eos}, we find 
\begin{equation}
\begin{split}
  s_r^{\,\,2} &= s_1^{\,\,2} = 0 \\
  s_t^{\,\,2} &= s_2^{\,\,2} = s_3^{\,\,2} = \tsfrac16(z^{-1}-z)^2.
\end{split}
\end{equation}
\begin{equation}
  s^2 = \tsfrac12\sum_{\mu=1}^3s_\mu^{\,\,2} = s_t^{\,\,2}. 
\end{equation}
With $\chi_1 = 0$ identically vanishing, we denote the only
independent $\chi_\mu$ by $\chi$ (without a basis index),
\begin{equation}
  \chi = -\chi_2 = \chi_3 = \tsfrac16(z^{-2}-z^2). 
\end{equation}
The anisotropy pressure scalar $q$ is now simply given by 
\begin{equation}\lbeq{qdef}
  q = \check\mu\chi. 
\end{equation}
To conclude this section we give the principal speeds for this
equation of state,
\begin{equation}\lbeq{radlongv}
\begin{split}
  v_{r||}^{\,2} &= \frac{\beta + 4[\sigma -
  (\check\Omega-\frac12)q]}{\rho+p_r} \\
  v_{t||}^{\,2} &= \frac{\beta + 2[\sigma +
  (\check\Omega-\frac12)q]}{\rho+p_t} \\
  v_{r\perp}^{\,2} &= \frac{\check\mu + \frac32(\sigma +
  q)}{\rho+p_t}   \\
  v_{t\perp r}^{\,2} &= \frac{ \check\mu +
  \frac32(\sigma-q)}{\rho+p_r} \\
  v_{t\perp t}^{\,2} &= \frac{\ts \check\mu + \sigma}{\rho+p_t},
\end{split}
\end{equation}
where, as in the general case, 
\begin{equation}
  \beta = \check\beta + \tsfrac43\check\mu + \left[
  (\check\Omega-1)\check\Omega +
  \check\beta\,\frac{d\check\Omega}{d\check{p}} \right] \sigma
\end{equation}


\section{Application to static spherically symmetric configurations}\label{sec:sss}
In this section we shall apply the elasticity theory just described to
study static spherically symmetric configurations, with neutron stars
as the obvious astrophysical objects in mind. It is remarkable that
although elastic spherically symmetric configurations have been
studied in the past \cite{mk:nonrot, park:elastsss}, it seems that not a
single solution has been published, whether exact or numerical, that can be
considered as a rough model for a neutron star. Referring to section
\ref{sec:deg} and using Schwarzschild coordinates, we write the
spacetime metric as
\begin{equation}
  g_{ab} = -u_a u_b + r_a r_b + t_{ab},
\end{equation}
where
\begin{align}
  u_a &= -e^\nu(dt)_a \\
  r_a &= e^\lambda(dr)_a, \quad e^{-2\lambda} = 1-\frac{2m}{r}  \\
  t_{ab} &= r^2(d\Omega^2)_{ab}, \quad d\Omega^2 = d\theta^2+\sin^2\!\theta\,d\phi^2. 
\end{align}
Generically, the stress-energy tensor will only be compatible with a
spherically symmetric spacetime if the material space metric $k_{AB}$
is also spherically symmetric and, in addition, the SO(3) symmetry
orbits of $M'$ and $X'$ are identified by the material space mapping
$\psi$. Hence, we set
\begin{equation} 
k_{AB}  = \tilde{r}_A\tilde{r}_B + \tilde{t}_{AB},
\end{equation}
where, analogously to the above, 
\begin{align}
  \tilde{r}_A &= e^{\tilde\lambda} (d\tilde{r})_A \\
  \tilde{t}_{AB} &=
  \tilde{r}^2(d\tilde\Omega^2)_{AB}, \quad d\tilde\Omega^2 =
  d\tilde\theta^2+\sin^2\!\tilde\theta\,d\tilde\phi^2.
\end{align}
The mapping $\psi$ is now, up to irrelevant SO(3) transformations,
defined through
\begin{equation}
  \tilde{r} = \tilde{r}(r), \quad d\tilde\Omega^2 = d\Omega^2.
\end{equation}
The pulled back material tensor $k_{ab}$ is given by eq.\ \refeq{kdeg} with
\begin{equation}\lbeq{nrntsss}
  n_r = e^{\tilde\lambda-\lambda}\,\frac{d\tilde{r}}{dr}, \quad n_t = \frac{\tilde{r}}{r}.
\end{equation}
It is common to assume a flat material space metric, \ie to set
$\tilde\lambda=0$. However, we will keep $\tilde\lambda$ unspecified
for the time being as other choices may be of interest as well\cite{cq:deform}. 
The particle density can be written
\begin{equation}\lbeq{neq}
n = n_r n_t^{\,2} = (\tilde{r}/r)^3z
\end{equation}
where, as in section \ref{sec:deg}, $z$ is the dimensionless quotient
\begin{equation}\lbeq{zeq}
  z = \frac{n_r}{n_t} = e^{\tilde\lambda-\lambda}\,\frac{r}{\tilde{r}}\frac{d \tilde{r}}{dr} 
\end{equation}
Regardless of the matter sources, the Einstein equations for any
static spherically symmetric spacetime can be formulated as (\cf \cite{bl:aniso})
\begin{align}\lbeq{nuevol}
  &\frac{d\nu}{dr} = \frac{m+\ts\frac12\kappa r^3 p_r}{r(r-2m)} \\ \lbeq{mevol}
  &\frac{dm}{dr} = \ts\frac12\kappa r^2\rho \\ \lbeq{prevol} 
  &\frac{dp_r}{dr} = -(p_r+\rho)\frac{m+\ts\frac12\kappa 
   r^3p_r}{r(r-2m)} + \frac{6q}{r}, \quad q = \ts\frac13(p_t-p_r) 
\end{align}
However, unless a one-parameter equation of state gives $\rho$ and
$p_t$ as known functions of $p_r$, this system is underdetermined. In
our case we are, as discussed in section
\ref{sec:deg}, effectively dealing with a two-parameter equation of state.  
An additional equation of motion can readily be found if we, to begin
with, use $n_r$ and $n_t$ as the two independent matter
variables. Indeed, eq.\ \refeq{nrntsss} directly implies that the
evolution equation for $n_t$ is
\begin{equation}\lbeq{ntevol}
  \frac{dn_t}{dr} =
  \frac1{r}\left(e^{\lambda-\tilde\lambda}n_r-n_t\right). 
\end{equation}
An evolution equation for $n_r$ can now be found by using
\begin{equation}
  \frac{dp_r}{dr} = \frac{\partial p_r}{\partial n_r}\frac{dn_r}{dr} +
  \frac{\partial p_r}{\partial n_t}\frac{dn_t}{dr},
\end{equation}
which we solve for $dn_r/dr$, yielding
\begin{equation}\lbeq{nrevol}
  \frac{dn_r}{dr} = \frac{n_r}{\beta_r}\left(\frac{dp_r}{dr} - \frac{\partial
  p_r}{\partial n_t}\frac{dn_t}{dr}\right),
\end{equation}
with $dp_r/dr$ and $dn_t/dr$ given by eq.\ \refeq{prevol} and eq.\ 
\refeq{ntevol}, respectively.  Knowing $\rho$, $p_r$ and $q$ as
functions of $n_r$ and $n_t$, eqs.\ \refeq{mevol}, \refeq{ntevol} and
\refeq{nrevol} provides a closed system of three equations for three
unknown variables $m$, $n_r$ and $n_t$. The gravitational potential
$\nu$ is decoupled from this system and hence its governing equation
\refeq{nuevol} can be integrated afterwards.

From a physical point of view, it is more natural to give the equation
of state in terms of $n$ and $z$ rather than in terms of $n_r$ and
$n_t$. To transform eqs.\
\refeq{ntevol} and \refeq{nrevol} into evolution equations for $n$ and
$z$, we begin by noting that $n_t^{\,3} = n/z$ leads to
\begin{equation}\lbeq{zpreevol}
  \frac{dz}{dr} = z\left(\frac1{n}\frac{dn}{dr} -
  \frac3{n_t}\frac{dn_t}{dr}\right),
\end{equation}
where $n_t^{-1}dn_t/dr$ can be reexpressed as
\begin{equation}\lbeq{ntevolnz}
  \frac1{n_t}\frac{dn_t}{dr} =
  \frac1{r}\left(e^{\lambda-\tilde\lambda}z-1\right).  
\end{equation}
Moreover, by viewing $p_r$ as a function of $(n,z)$ and using eq.\
\refeq{zpreevol}, we find
\begin{equation}
  \frac{dp_r}{dr} = \frac{\beta_r}{n} \frac{dn}{dr} -
  \frac3{n_t}\frac{dn_t}{dr}\,z\frac{\partial p_r}{\partial z},
\end{equation}
which can be solved for $dn/dr$, resulting in
\begin{equation}\lbeq{nevol}
  \frac{dn}{dr} =
  \frac{n}{r\beta_r}
  \left[-(\rho+p_r)\frac{m+\ts\frac12\kappa
  r^3p_r}{r-2m} + 6q + 3z\frac{\partial p_r}{\partial z}
  \left(e^{\lambda-\tilde\lambda}z-1\right)\right], 
\end{equation}
where we have used eqs.\ \refeq{prevol} and
\refeq{ntevolnz}. Inserting eq. \refeq{ntevolnz} into eq.\
\refeq{zpreevol} as well, the evolution equation for $z$ is
\begin{equation}
  \frac{dz}{dr} = z\left[\frac1{n}\frac{dn}{dr} -
  \frac3{r}\left(e^{\lambda-\tilde\lambda}z-1\right)\right],
\end{equation}
where $dn/dr$ is to be inserted from eq.\ \refeq{nevol}. Summarising,
in the variables $(m,n,z)$, Einstein's equations take the form
\begin{align}\lbeq{dmdrpre}
  &\frac{dm}{dr} = \ts\frac12\kappa r^2\rho \\ \lbeq{dndrpre}
  &\frac{dn}{dr} =
  \frac{n}{r\beta_r}
  \left[-(\rho+p_r)\frac{m+\ts\frac12\kappa
  r^3p_r}{r-2m} + 6q + 3z\frac{\partial p_r}{\partial z}
  \left(e^{\lambda-\tilde\lambda}z-1\right)\right] \\ \lbeq{dzdrpre}
  &\frac{dz}{dr} = z\left[\frac1{n}\frac{dn}{dr} -
  \frac3{r}\left(e^{\lambda-\tilde\lambda}z-1\right)\right],
\end{align}
where $\rho$, $p_r$, $q$ and $\beta_r$ are related by a two-parameter
equation of state, given by $\epsilon = \epsilon_{\mathrm{eff}}(n,z)$
and
\begin{equation}
  \rho = n\epsilon, \quad p_r = n^2\frac{\partial\epsilon}{\partial n}
  - 2q, \quad q = -{\ts\frac12}nz\frac{\partial\epsilon}{\partial z},
  \quad \beta_r = n\frac{\partial p_r}{\partial n} +
  z\frac{\partial p_r}{\partial z}.  
\end{equation}
We shall now specialise the system of equations \refeq{dmdrpre} -
\refeq{dzdrpre} to the particular equation of state introduced in
section \ref{sec:eos}. To facilitate a comparison with the perfect
fluid case, we use the relation $dn/n=d\check{p}/\check\beta$ to
replace the particle density $n$ with the unsheared pressure
$\check{p}$ as one of the independent variables. We find that, with
$(m,\check{p},z)$ as independent variables, Einstein's equations can
be put in the form
\begin{align}
    &\frac{dm}{dr} = \ts\frac12\kappa r^2\rho \\ 
    &\frac{d\check{p}}{dr} =
    \frac{\check\beta}{r\beta_r}\left\{
    -(\rho+p_r)\frac{m+\ts\frac12\kappa
    r^3p_r}{r-2m} + 6q + 4\left(e^{\lambda-\tilde\lambda}\,z - 
    1\right)\left[ \check\mu + 3\sigma + \tsfrac32(1-\check\Omega)q \right]
    \right\} \\
    &\frac{dz}{dr} =
    \frac{z}{r}\left[\frac{r}{\check\beta}\frac{d\check p}{dr} -
    3\left(e^{\lambda-\tilde\lambda}\,z - 1\right)
    \right],  
\end{align}
where 
\begin{align}
  &\sigma = \check\mu s^2, \quad s^2 = \tsfrac16(z^{-1}-z)^2, \\
  &q = \check\mu\chi, \quad \chi = \tsfrac16(z^{-2}-z^2), \\ 
  &\rho = \check\rho + \sigma, \\
  &p_r = p - 2q, \quad p = \check{p} + (\check\Omega-1)\sigma,
  \lbeq{prad} \\ 
  &\check\beta =
  (\check\rho+\check{p})\frac{d\check{p}}{d\check\rho}, \\
  &\beta_r = \beta + 4\left[\sigma + (\check\Omega-\tsfrac12)q\right],
  \quad \beta = \check\beta + \tsfrac43\check\mu + \left[
    \check\Omega(\check\Omega-1) +
    \check\beta\,\frac{d\check\Omega}{d\check{p}} \right] \sigma, \\ 
  &\check\Omega =
  \frac{\check\beta}{\check\mu}\frac{d\check\mu}{d\check{p}}.
\end{align}
and where $\check\rho$ and $\check\mu$ are related to $\check{p}$
through the equation of state. Note that, for vanishing shear modulus
$\check\mu$, the equation for $z$ is decoupled and the remaining
system reduces to the standard perfect fluid form. This makes this
explicitly closed system useful for comparing elastic models
with the corresponding perfect fluid ones.

It is interesting to evaluate the derivative $dz/dr$ at zero shear
$z=1$. Using the fact that $\sigma = q = 0$ at such a point we readily
find
\begin{equation} 
\left.\frac{dz}{dr}\right|_{z=1} = -\frac1{r v_{||}^{\,2}}\left[\frac{m + 
\tsfrac12\kappa r^3\check{p}}{r-2m} + 3\frac{d\check{p}}{d\check\rho}
\left(e^{\lambda-\tilde\lambda}-1\right)\right] 
\end{equation}
Using a flat material space metric, \ie setting $\tilde\lambda=0$, the
right hand side is clearly negative for any reasonable equation of
state, since $\lambda$ is then everywhere
positive\cite{mms:nohorizon}. Therefore, unless one chooses $z>1$ at a
phase transition, see below, $z$ will always remain less than or equal
to unity. Hence the radial pressure can never exceed the tangential
pressure unless one chooses it to do so at the phase transition. One
may suspect that such a choice of phase transition would tend to make
the star radially unstable, but this will be the subject for later
investigation\cite{ks:stability}.


\section{Boundary conditions}\label{sec:bc}

There are three different types of boundaries that one might consider;
the center of the star $r=0$, a transition between different materials
in the star and the surface of the star $r=R$. At the center we
require that the spacetime is regular (elementary flat).  This implies
pressure isotropy $p_{ab}=p\lagom h_{ab}$ and, given this, the
conditions are the same as for a perfect fluid, namely finiteness of
the energy density $\rho$ and pressure $p$ as well as vanishing of the
mass function $m$. Since the pressure isotropy condition is fulfilled
iff $q=0$, eq.\ \refeq{qdef} shows that unless $\check\mu = 0$, the
function $z$ must take its unsheared value $1$, meaning that the shear
$s^2$ must vanish. In both cases $\check{p}=p_r=p_t=p$ and
$\rho=\check\rho$ hold at the center.  Since the right hand sides of
the evolution equations are all zero at $r=0$ we are forced to move
out the starting point of integration a small distance from the center
when solving the evolution equations numerically. The initial
conditions for the integration is thus given by the following
expansions around $r=0$:
\begin{align}
  &m = \tsfrac16\kappa\check\rho_c\,r^3 + \ldots \\
  &\check{p} = \check{p}_c -
  \kappa\,\check\beta_c\frac{(\check\rho_c+\check{p}_c)(\check\rho_c + 3\check{p}_c)
  - 4\check\mu_c\check\rho_c}{12\left( \check\beta_c +
  \tsfrac43\check\mu_c \right)}\,r^2 + \ldots \\  
  &z = 1 - \kappa\frac{(\check\rho_c + \check{p}_c)(\check\rho_c
  + 3\check{p}_c) +
  3\check\beta_c\check\rho_c}{30\left( \check\beta_c +
  \tsfrac43\check\mu_c \right)}\,r^2 + \ldots 
\end{align}
where a subscript $c$ denotes evaluation at $r=0$.

The remaining types of boundary conditions are governed by the
continuity of the first and second fundamental
forms\cite{israel:junction}, which for all static spherically
symmetric spacetimes leads to continuity of the Schwarzschild radial
coordinate $r$, the mass function $m$ (no surface layers), the
gravitational potential $\nu$ and its first derivative $d\nu/dr$. Via
eq.\
\refeq{nuevol}, the continuity of $d\nu/dr$ can be replaced by continuity
of the radial pressure $p_r$. Since we are using $r$ as our radial
variable, its continuity requirement is trivially imposed and will not
be discussed further.

At the surface of the star where we match the interior solution to a
Schwarzschild vacuum solution of mass $M$ the boundary conditions
reduce to
\begin{align}
  &m|_{r=R} = M \\
  &\nu|_{r=R} = \frac12\ln{\left(1-\frac{2M}{R}\right)} \\
  &p_r|_{r=R} = 0.  \lbeq{psurf}
\end{align}
An interesting observation is that, for $\check\mu = \muk\check p$
with $\muk$ constant, eqs.\ \refeq{psurf} and \refeq{prad} imply
\begin{equation}
  z^2|_{r=R} = \frac{-3-(1-\check\Omega_s)\muk+\sqrt{9+6(1-\check\Omega_s)\muk
  +4\muk^2}}{\muk(1+\check\Omega_s)}
\end{equation}
where $\check\Omega_s=\check\Omega|_{r=R}$. Remarkably, if the
isotropic part of the equation of state is a relativistic polytrope,
in which case $\check\Omega$ is constant (and actually equal to the
adiabatic index, see section \ref{sec:model}), it follows that the
value of $z$ at the surface is known
\emph{a priori} in terms of the equation of state parameters $\check\Omega$ 
and $\muk$.

Turning now to an interior boundary between two different phases of
matter, the continuity of $m$ and $\nu$ can be imposed without
difficulties. On the other hand the continuity of the radial pressure
$p_r$ requires some thought since we are using the unsheared pressure
$\check{p}$, rather than $p_r$, as an unknown variable. Since the
junction conditions have nothing directly to say about the shear
variable $z$, there is in fact no unique way of making the matching.
This means that the uniqueness result of Park\cite{park:elastsss} does
not generalise to the case of elastic models with transitions between
different phases of matter; if phase transitions are allowed for there
is \emph{not} a unique static spherically symmetric model for a given
value of the central pressure $p_c$. Since a real neutron star hardly
can be viewed as one big piece of elastic matter, this complication is
not of pure academic nature. Although a particular class of equations
of state is considered here, it should be obvious that non-uniqueness
is not peculiar to this class. However, we shall see below that the
case when the outer matter phase is a perfect fluid is an important
exception for which uniqueness of the matching does apply.

Now, according to eq.\ \refeq{prad}, we are to match the following
combination of the unknown variables $\check{p}$ and $z$:
\begin{equation}
  \check{p} - \frac{\check\mu}6\left[(1-\check\Omega)(z^{-1}-z)^2+2(z^{-2}-z^2)\right],
\end{equation}
where it is to be recalled that $\check\mu$ and $\check\Omega$ are in
general functions of $\check{p}$. For simplicity as well as for all
practical purposes of this paper, we shall restrict ourselves to the
situation when one of the phases is a perfect fluid ($\check\mu$ identically
vanishing) while the other is a general elastic phase.  Since the
equations are always integrated from the center and outwards, the
values $\check{p}_-$ and $z_-$ of $\check{p}$ and $z$ on the interior side
of the boundary will be given while the exterior values $\check{p}_+$
and $z_+$ are to be determined. In fact, this implies that the cases
when the elastic phase is interior to the perfect fluid phase and vice
versa need to be treated separately. Before we do so, we note that for
the perfect fluid phase, the value of $z$ is completely irrelevant so
in fact it is meaningless to even define $z$ in such a region of the
star. Now, starting with an elastic phase interior to a perfect fluid,
the matching condition becomes
\begin{equation}
 \check{p}_+ = \check{p}_-  -  \frac{\check\mu_-}6\left[(1-\check\Omega_-)(z_-^{\,-1}-z_-)^2+2(z_-^{\,-2}-z_-^{\,2})\right],
\end{equation}
which uniquely determines $\check{p}_+$, while, as we just concluded,
$z_+$ need not even be defined. If, on the other hand, the interior
phase is the perfect fluid, the junction condition is
\begin{equation}
  \check{p}_+  -  \frac{\check\mu_+}6\left[(1-\check\Omega_+)(z_+^{\,-1}-z_+)^2+2(z_+^{\,-2}-z_+^{\,2})\right] = \check{p}_-.
\end{equation}
This leads to a one parameter family of possible phase transitions
characterised for instance by the value of $z_+$. Setting $z_+=1$, it follows that
$\check{p}$ as well as all other pressures are continuous across the
boundary.\footnote{Note however that the total energy density $\rho$
may still be discontinuous, but only if $\check\rho$ is.} This is
clearly the simplest case to consider.  It is interesting to observe
that although $z_+=1$ implies continuity of $\check{p}$ the converse is
not necessarily true. Namely, setting $\check{p}_+ = \check{p}_-$ leads to
two solutions for $ z_+^{\,2}$,
\begin{equation}
 z_+^{\,2} = 1 \quad \textrm{or}\quad  z_+^{\,2} = -\frac{3-\check\Omega_+}{1+\check\Omega_+}, 
\end{equation}
where the second equality can only be valid if $\check\Omega_+ >
3$. The second solution does not imply continuity of the other
pressures but is probably unphysical due to the large value of $\check\Omega_+$.

\section{Example: relativistic polytropes}\label{sec:model}%
In order to concretise the theory developed so far, we shall here
consider a few examples of static spherically symmetric
configurations. The first step is to specify the equation of state in
terms of two functions of state \eg $\check\rho(\check{p})$ and
$\check\mu(\check{p})$. For simplicity we shall only consider models
where the one-parameter relation between $\check\rho$ and $\check{p}$
is that of a relativistic polytrope, \ie
\begin{equation}\lbeq{eos}
  \check{\rho} = \frac{p_1}{\check\Gamma-1}
  \left[\frac{\check{p}}{p_1} +
  \left(\frac{\check{p}}{p_1}\right)^{\!\!1/\check\Gamma}\,\right], 
\end{equation}
where $\check\Gamma = \check\beta/\check{p}$ is the (constant)
\emph{adiabatic} or
\emph{compressibility index} and $p_1$ is the pressure which signals
the transition between the classical polytropic
$\check{\rho}\propto\check{p}^{\check\Gamma}$ and the linear
$\check{\rho}\propto\check{p}$ regime. More precisely, at $\check{p} =
p_1$, the polytropic part and the linear part contributes equally to
$\check\rho$. This unsheared part of the equation of state follows
from
\begin{equation}
  \check{\epsilon} = \frac{p_1 \dumkonstant}{\check\Gamma-1}\left[(\dumkonstant n)^{\check\Gamma-1}+1\right]
\end{equation}
where $\dumkonstant$ is some constant with the dimension of volume. The shear
modulus $\check\mu$ will be assumed to be piecewise proportional to
the unsheared pressure, $\check{\mu} = \muk\check{p}$, where $\muk$ is
a dimensionless constant which is allowed to vanish. This assumption
is supported by several calculations regarding the equation of state
of neutron stars, see \eg \cite{haensel:solid} for a review. For this
class of equations of state the compressibility index $\check\Gamma$
will be constant throughout the star.  Moreover $\check\Omega$ will be
piecewise constant and equal to $\check\Gamma$ where it is non-zero.
We chose the constant to be $\check\Gamma=5/3$ corresponding to a
non-relativistic Fermi gas. The transition pressure $p_1$ is chosen to
be $1.5\times10^{37}$ dyn/cm${}^2$ in order to give total radii in the
order of 10 km. We shall also use the simplifying
assumption that the material space metric is flat, \ie we choose
$\tilde\lambda=0$.

We consider first the situation where we have a rigid crust afloat on
a perfect fluid core. The transition from the fluid to the solid phase
is taken to be approximately at the neutron drip pressure,
$\check{p}=6\times10^{29}$ dyn/cm${}^2$, and we let $\muk$ take
values in the range $0\le\muk\le0.1$. Of course a zero value
corresponds to a perfect fluid star and therefore gives a good
reference model. The highest value 0.1 is probably unphysically large
but is included in order to give insights to extreme cases. The phase
transition is chosen such that all pressures are continuous across the
matching surface. The first question we ask is how the maximum mass
and stability properties change when we include the effect of a
crust. For a few models with different $\muk$ we have plotted the
total mass $M$ as a function of the total radius $R$. This allows us
to read off the maximum mass attainable for these models. Moreover, at
least in the case of a perfect fluid, an extremum of this curve
correspond to a change in the stability for one radial normal
mode\cite{thorne:struct}. Whether or not this is also the case when
the matter is allowed to behave elastically is currently being
investigated\cite{ks:stability}, but we anticipate that it will due to
the following argument. For well-behaved equations of state the
problem of calculating time-harmonic radial perturbations turn into a
Sturm-Liouville problem with the squared frequency $\omega^2$ as the
eigenvalues. Since the eigenvalues of such a problem can be ordered,
with the lowest eigenvalue $\omega_0^2$ corresponding to an
eigenfunction without nodes, it is clear that instability sets in at
points where $\omega_0=0$. Allowing only perturbations which conserves
the total mass of the star it is seen that only stationary points in
the mass-radius diagram correspond to zero frequency modes, since it
is only there the perturbed star is still in the equilibrium
sequence, and hence does not oscillate. We point out however that if
the equation of state involves discontinuities (as, for instance, is
the case for our choice of shear modulus) one has to be careful since
the perturbation problem will in general no longer be of the strict
Sturm-Liouville type. For the types of discontinuities considered in
this paper we do not expect that the general conclusion will change.

\begin{figure}[ht]
 \centering \begin{minipage}[t]{0.78\linewidth}
  \centering
  \includegraphics[width=0.7\textwidth, angle=270]{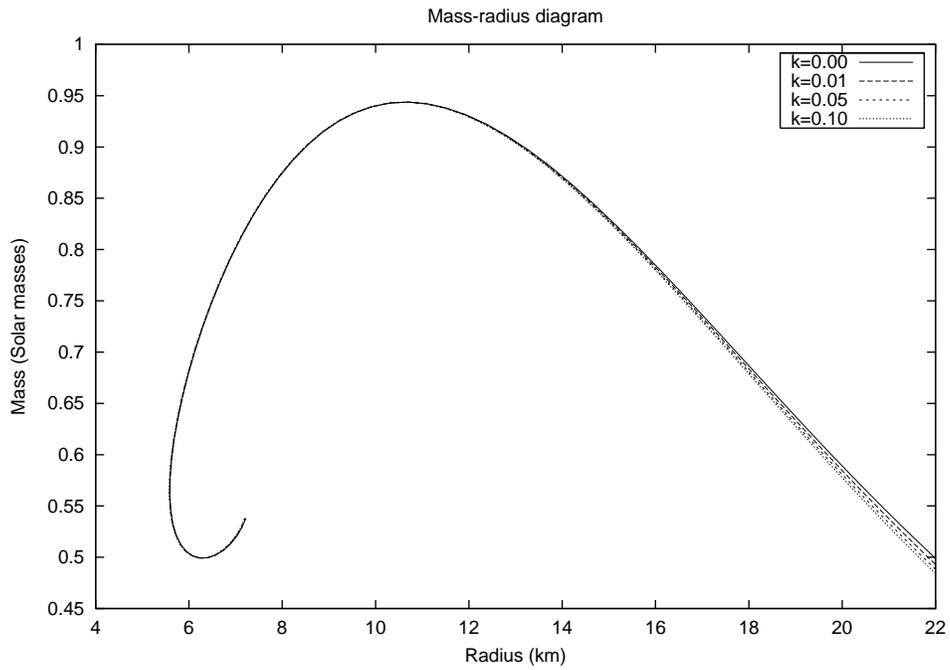}
   \caption{Mass-radius diagram for the models with a solid crust. }
 \label{fig:crust1}
 \end{minipage}
\end{figure}

\begin{figure}[hb]
 \centering\begin{minipage}[t]{0.78\linewidth} \centering
 \includegraphics[width=0.7\textwidth, angle=270]{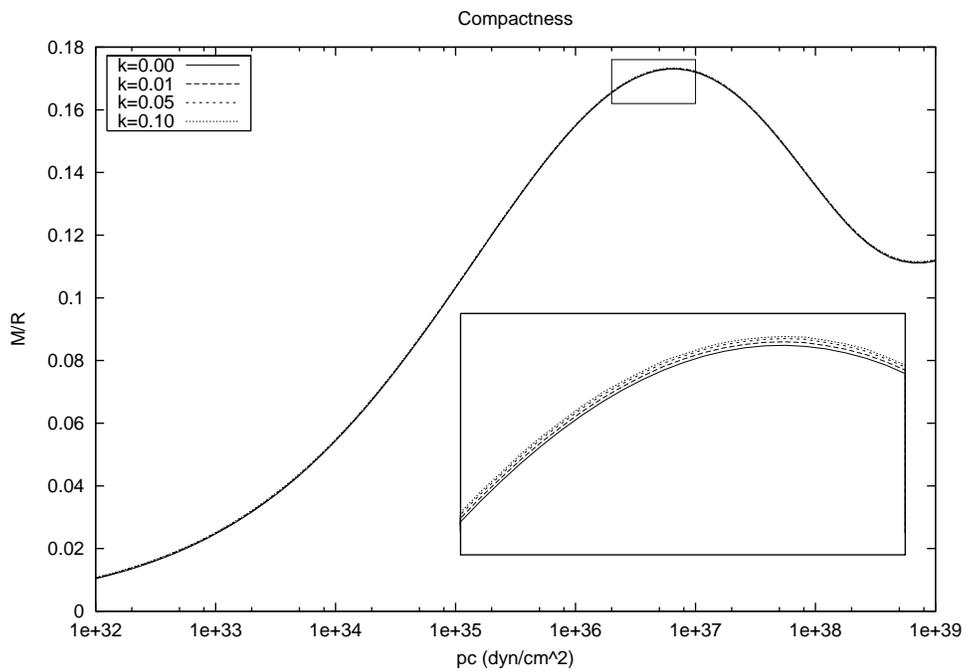}
 \caption{Compactness as a function of central pressure for models
 with a solid crust. Inserted is a magnification of a portion of the
 graph. There it can be seen that the compactness increases with
 increasing shear modulus, although the effect is very small.} 
\label{fig:crust2} \end{minipage}
\end{figure}

In figure \ref{fig:crust1} we see, as expected, that adding a solid
crust does not affect the maximum mass significantly, even for very
large $\muk$. However, for stars with large total radii, corresponding
to low central pressures, the curves differ significantly. One should
however be careful when interpreting this diagram since each model
should be viewed as parametrised by the central pressure which is not
apparent in fig.\ \ref{fig:crust1}. For this reason it is useful to
plot also the \emph{compactness} $M/R$ as a function of the central
pressure, as done in fig.\  \ref{fig:crust2}. Here it is clear that
the compactness does not change significantly with increasing shear
modulus, even for low central pressures. However, the obvious trend is
that the compactness increases with increasing shear modulus.

\begin{figure}[ht]
  \centering\begin{minipage}[t]{0.78\linewidth}
 \centering
 \includegraphics[width=0.7\textwidth,angle=270]
 {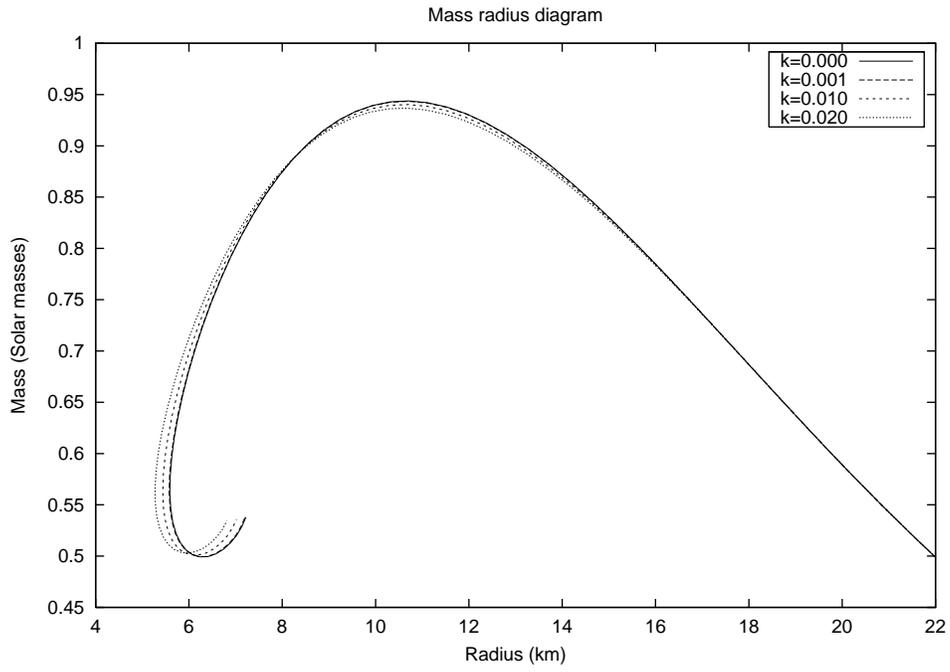}
 \caption{Mass radius diagram for models with a solid core.}
 \label{fig:core1} 
  \end{minipage} 
 \end{figure}

We turn, now, to the case when we instead have a solid matter phase in
the core. We take the transition pressure to be $2\times10^{34}$
dyn/cm${}^2$. In this case the proportionality constant $\muk$ is
expected to be smaller than in the crust\cite{haensel:solid}, if it is
non-zero at all. We therefore choose to look at the interval $0 \le
\muk \le 0.02$, where again the largest value is expected to be
unphysical. We again look at the mass-radius diagram (figure
\ref{fig:core1}) and see that the maximum mass actually
\emph{decreases} with increasing $\muk$ in this case. However, 
looking at the compactness we note that the maximum value
increases. This can be understood from the mass-radius diagram since
the compactness can be parametrised by the angle between a ray from
the origin to $(r,m)$ and the $r$-axis. Clearly the maximum
compactness is reached after the peak of the mass-radius curve,
corresponding to an unstable model.  One interesting possibility for
neutron stars is that they, if compact enough, could allow trapped
$w$-modes\cite{cf:gwresonance}. Based on perfect fluid calculations
however, it does not seem likely that this is the case for real
neutron stars\cite{ivd:ultracompact}. Speculatively, adding a rigid
core could provide the extra compactness needed. As discussed by
Rosquist\cite{rosquist:trapped}, it is not the total compactness that
is important but rather some partial measure of compactness that
determines if a stellar configuration allow for trapped $w$-modes. A
necessary (but not sufficient) condition is that $\nu'r=1$ somewhere
in the star. The models here does not allow for trapped $w$-modes, but
a higher $\muk$ in the core does indeed pushes $\nu'r$ towards one.

In order to assess the impact of the choice of parameters we have
calculated a large number of models with various parameter values. The
effects of a non-zero shear modulus do not appear to depend strongly
on the choice of $p_1$ or $p_t$. Also, adding even rather
large discontinuities in the tangential pressure in the fluid-solid
phase transition does not seem to have any major impact either.


\section{Comments and outlook}\label{sec:conclude}%

As far as our study of static spherically symmetric solutions to the
Einstein equations with elastic matter source is concerned, we see it
as a first foundation laying step towards a study of more generic
configurations. The obtained numerical solutions may not be highly
interesting in themselves, as they, at least for realistic choices of
the shear modulus $\check\mu$, do not differ drastically from the
corresponding perfect fluid model. However, it is important to have a
good control and understanding of the background model when perturbing
a spherically symmetric solution into less symmetric neighbours. A
feature of the presented models which we see as oversimplified is that
the equation of state is set up using a single flat material space
metric throughout a large region of the star, stretching over several
orders of magnitude in energy density and pressure. While the usage of
a flat material space metric is naturally suggested from first
principles (see the end of section \ref{sec:eom}) one should not
expect the stellar material to be ordered over such long ranges as
being considered here, which means that the use of a flat metric as
the single reference structure is not well justified. Instead, the
crust of a neutron star is layered where each layer has a different
nuclear composition and lattice structure\cite{haensel:solid}. Thus, a
proper neutron star model should include several phases of matter,
each well, at least effectively, described by a flat material space
metric. We believe that our formalism is well suited to deal with
these kinds of problems as well as with perturbation calculations of
various kinds.

\subsection*{Acknowledgements}
We are grateful to prof. Kjell Rosquist and to Moundheur Zarroug for
helpful discussions. We would also like to thank the anonymous
referees for pointing out references to earlier work that were
initially overlooked by us.

\section*{Appendix A: Expansion around a locally unsheared state}

To motivate our choice of invariant as well as our equation of state
it is instructive to make an expansion of the energy around a locally
unsheared state, \ie { }a state where $h_{ab} = n_0^{-2/3}k_{ab}$
for some particle density $n=n_0$. Using the notation
$\gtnoll_{ab} =\eta_{ab}|_{n=n_0} =
n_0^{-2/3}k_{ab}$, we introduce the \emph{strain tensor} $e_{ab}$ as
\begin{equation}
  e_{ab} = \tsfrac12(h_{ab}-\gtnoll_{ab}). 
\end{equation}
In the following, we shall work exclusively with tensors having mixed
indices and formulate everything as (three-dimensional) matrix equations in index free
notation. Thus needing to raise one of the indices of $e_{ab}$, we
have two natural choices to do so, either by using $g^{ab}$ or
$\gtnoll^{-1\,ab} = n_0^{2/3}k^{-1\,ab}$. We shall for the moment work
with both, as the former is more natural from a spacetime point of
view, while the latter connects more closely to the classical theory and
is the one used by Carter and Quintana.

Now, with these conventions we introduce $\mathbf{e}$ and
$\mathbf{\hat{e}}$ to be the matrices with indices
\begin{equation}
e^a{}_b = \ts\frac12 (h^a{}_b - \gtnoll^{a}{}_b),  \quad 
\hat{e}^a{}_b = \ts\frac12 (\gtnoll^{-1\, a}{}_b - h^a{}_b),  
\end{equation}
or in index-free form
\begin{equation}
\mathbf{e} = \tsfrac12 (\mathbbm{1} - \bgtnoll), \quad 
\mathbf{\hat{e}} = \tsfrac12 (\bgtnoll^{-1} - \mathbbm{1}). 
\end{equation}
Note that, formally, $\mathbf{e}$ and $\mathbf{\hat{e}}$ are related by
\begin{equation}
  \mathbf{\hat{e}} = \frac{\mathbf{e}}{\mathbbm{1}-2\mathbf{e}}, \quad  \mathbf{e} = 
  \frac{\mathbf{\hat{e}}}{\mathbbm{1}+2\mathbf{\hat{e}}}.
\end{equation}
To proceed we expand the particle density to second order in
$\mathbf{e}$ and $\mathbf{\hat{e}}$ respectively,
\begin{align}\lbeq{novern0}
  \frac{n}{n_0} &\simeq 1-\Tr{\mathbf{e}}+\tsfrac12(\Tr{\mathbf{e}})^2-\Tr{\mathbf{e}^2}  \\
    &\simeq 1-\Tr{\mathbf{\hat{e}}}+\tsfrac12(\Tr{\mathbf{\hat{e}}})^2+\Tr{\mathbf{\hat{e}}^2},
\end{align}
where ``$\,\simeq\,$'' means ``equal up to and including quadratic
terms in $\mathbf{e}$''. Note that the only formal difference between
these two expressions is the plus sign in front of
$\Tr{\mathbf{\hat{e}}^2}$. So, since any quantity expanded to second
order in $\mathbf{e}$ will be a linear function of $\Tr{\mathbf{e}}$,
$\Tr{\mathbf{e}^2}$ and $(\Tr{\mathbf{e}})^2$, it follows that such an
expansion will look formally the same when instead expressed in
$\mathbf{\hat{e}}$, if we use eq.\
\refeq{novern0} to replace all occurrences of $\Tr{\mathbf{\hat{e}}}$
in favour of $n/n_0-1$. In the following we will therefore use only
$\mathbf{e}$.  Now, in the classical theory of elasticity, one usually
relates thermodynamical quantities to unit volume of the completely
relaxed reference state.  Hence, if we as our reference state take the
unsheared state $\mathbf{e} = 0$, which however is not necessarily
completely relaxed, it is $E=n_0\epsilon$ rather than $\rho =
n\epsilon$ that should be compared to the energy per unit volume of
the classical works.  Without assuming anything about the equation of
state we write the expansion as
\begin{equation}\lbeq{Free_exp}
  E \simeq E_0 +
  p_0\left(\frac{n}{n_0}-1\right)+\ts\frac12\lambda_0(\Tr
  \mathbf{e})^2 + \mu_0\Tr(\mathbf{e}^2),
\end{equation}
where $E_0$, $p_0$, $\lambda_0$ and $\mu_0$ are the expansion
coefficients. In general, these coefficients depend on the choice of
$n_0$.  It is apparent from eq.\ \refeq{Free_exp} that the unsheared
state can only be a completely relaxed state, \ie a state that
minimises $E$ and hence also $\epsilon$, if
\begin{equation}\lbeq{mincond}
  p_0 = 0, \quad \lambda_0 \ge 0, \quad \mu_0 \ge 0.
\end{equation}
In this case we may identify the constants $\lambda_0$ and $\mu_0 $ as
the \emph{Lam\'{e} coefficients}. The constant
$\mu_0$ is also called the \emph{shear modulus}, while the combination
$\beta_0 = \lambda_0 + \frac23\mu_0$ is called the \emph{bulk
  modulus}. However, we warn the reader that this definition of the
bulk modulus does not generalise to the case $p_0\neq 0$, a point
which we will return to later in this appendix.

We may find an interpretation for $p_0$ if we note that for small
$\mathbf{e}$ the linear term dominates over the quadratic and higher
order terms and we have approximately the thermodynamic relation
\begin{equation}
\Delta E  \approx  - p_0 \Delta V,
\end{equation} 
where $\Delta E=E-E_0$, $\Delta V = -(n-n_0)/n$ with $\Delta V$ being  the
change in the volume. Thus it follows that $p_0$ may be identified as pressure. In
situations with small pressure, for instance near the surface of a
star, it would be sufficient to use a \emph{Hookean}  equation of
state \ie to let the equation of state be given by eq.\ 
\refeq{Free_exp} with the cubic and higher order terms simply dropped
and with the conditions \refeq{mincond} satisfied.  However, since we
are interested in the interior of neutron stars where pressures can be
extreme, a more general approach is called for. As noted previously in
section \ref{sec:relasticity} we want the scalar invariants of
$\mathbf{k}$ used to set up the equation of state to consist of the
particle density $n$ along with up to two scalar invariants of the
unit determinant matrix $\pmb\eta$.  As the invariants
$\Tr{\mathbf{e}}$ and $\Tr(\mathbf{e}^2)$ of $\mathbf{e}$ used in the
above expansions are not exact invariants of $\pmb\eta$, we need to
reformulate things a little bit. So, following CQ we
introduce the \emph{constant volume shear tensor} $s_{ab}$
\begin{equation}
  s_{ab} = \tsfrac12(h_{ab}-\eta_{ab}),
\end{equation}
which in contrast to the shear tensor $e_{ab}$ does not depend on the
particle density $n_0$ of the reference state. As was the case with
the shear tensor, there are two natural ways to raise one index,
namely by using either $h^{ab}$ or $\eta^{-1\, ab}$, giving us the
matrices
\begin{equation}
  \mathbf{s} = \ts\frac12(\mathbbm{1} - \pmb{\eta}), \qquad
  \mathbf{\hat{s}} = \ts\frac12(\pmb{\eta}^{-1} -
  \mathbbm{1})
\end{equation} 
Expanding an arbitrary integer power of $\pmb\eta$ to second
order in $\mathbf{e}$ gives
\begin{equation} 
  \Tr{(\pmb{\eta}^n)} \simeq 3 + 2n^2s^2, \qquad n\in \mathbb{Z}, 
\end{equation}
where
\begin{equation} \lbeq{s2eq}
   s^2 \simeq \Tr{(\mathbf{e}^2)}-\tsfrac13(\Tr{\mathbf{e}})^2 =
   \Tr{\left\{ \left[ \mathbf{e}-\tsfrac13(\Tr{\mathbf{e}})\mathbbm{1}
   \right]^2 \right\}}.
\end{equation}
Note that $\pmb{\eta}$ has only one independent invariant up to second
order in $\mathbf{e}$, which therefore also will be the case for
$\mathbf{s}$. Namely, it follows that
\begin{equation}\lbeq{seq}
  \Tr{\mathbf{s}} \simeq \Tr{\mathbf{s}^2} \simeq s^2 \simeq
          \Tr{\mathbf{\hat{s}}} \simeq \Tr{\mathbf{\hat{s}}^2}.
\end{equation}
It is now instructive to reexpress the energy $E$ as a function of
$n/n_0-1$ and $s^2$. Using \refeq{novern0} we have, 
\begin{equation}\lbeq{Freens2}
  E \simeq E_0 + p_0\left(\frac{n}{n_0}-1\right) + \tsfrac12
  (\lambda_0+\tsfrac23\mu_0)\left(\frac{n}{n_0}-1\right)^{\!\!2} +
  \mu_0 s^2. 
\end{equation}
Due to the relations \refeq{seq}, which imply that $\Tr{\mathbf{s}}$
is quadratic in $\mathbf{e}$ to lowest order, it follows that up
second order in $\mathbf{e}$, the scalar $s^2$ is expressible, but
actually not \emph{uniquely} expressible, in terms of linear and
quadratic invariants of $\mathbf{s}$. Any function of the form $s^2
\simeq a \Tr \mathbf{s} +(1-a) \Tr \mathbf{s}^2 + b (\Tr
\mathbf{s})^2$ will give the same energy contribution up to
that order.  For the time being we simply assume that $s^2$ is any
such function and postpone a discussion of how to make an appropriate
choice of it until later.  To interpret eq.\
\refeq{Freens2} physically, we note that the first three terms, i.e.
\begin{equation}
   E_0 + p_0\left(\frac{n}{n_0}-1\right)
   +\tsfrac12(\lambda_0+\tsfrac23\mu_0)\left(\frac{n}{n_0}-1\right)^2  
\end{equation}
gives the compression part, or ``perfect fluid part'', of the equation
of state, while the last term $\mu_0s^2$ adds to the equation of state
a resistance against shear, \ie against matter deformations that are
not pure volume changes. The strength of this resistance is given by
the shear modulus $\mu_0$. To extend this local expansion of the
elastic matter equation of state into an equation of state valid for
all admissible particle densities $n$, without neither introducing a
third invariant nor inserting a nonlinear dependence on $s^2$, we
simply assume that the energy per particle $\epsilon$ is a first order
polynomial in $s^2$ with coefficients of arbitrary $n$-dependence.
Following our previous convention that quantities depending only on
$n$ are checked, we have
\begin{equation}
  \epsilon = \check\epsilon_0 + \check\epsilon_1s^2.  
\end{equation}
Since the expansion of $\epsilon$ to second order in $\mathbf{e}$ is
\begin{equation}
  \epsilon \simeq \check\epsilon_0(n_0) + \check\epsilon{\,}_{\!0}'(n_0)(n-n_0) + {\ts\frac12}
  \check\epsilon{\,}_{\!0}''(n_0)(n-n_0)^2 + \check\epsilon_1(n_0)s^2 
\end{equation}
we can immediately see that the values of the expansion coefficients
at $n=n_0$ are
\begin{align}
  E_0 &= n_0\check\epsilon_0 \\
  p_0 &= n_0^2 \check\epsilon{\,}_{\!0}'(n_0) \\
  \lambda_0  &= n_0^3\check\epsilon{\,}_{\!0}''(n_0) - \ts\frac23 \mu_0 \\
  \mu_0 &= n_0\check\epsilon_1(n_0) 
\end{align}
Since these relations hold for any $n_0$, we may introduce
$n$-dependent elasticity parameters according to
\begin{equation}
\begin{split}
  &\check\rho = n\check\epsilon_0 \\
  &\check p = n^2\check\epsilon{\,}_{\!0}'  \\
  &\check\lambda = n^3\check\epsilon{\,}_{\!0}'' - \tsfrac23\check\mu \\
  &\check\mu = n \check\epsilon_1.
\end{split}
\end{equation}
The parameters $\check\rho$, $\check{p}$ and $\check\mu$ will be
called the \emph{unsheared energy density}, the \emph{unsheared
pressure} and the \emph{shear modulus}, respectively. The parameter
$\check\lambda$, whose value at $\check{p}=0$ is one of the classical
Lam\'e coefficients, does not have a useful physical interpretation
and will not be mentioned by us elsewhere. Instead, the information
given by the second derivative of $\check\epsilon_0$ will be described
by the \emph{bulk modulus}
\begin{equation}
  \check\beta = n\frac{d\check{p}}{dn} = n^3\check\epsilon{\,}_{\!0}''
  + 2\check{p} = \check\lambda + 2\check{p} + \tsfrac23\check\mu. 
\end{equation}

Let us now discuss the choice of shear invariant $s^2$ which so far
only is defined up to second order in $\mathbf{e}$. The primary
requirements that we which to impose are:
\begin{itemize}
 \item{$s^2$ should locally, around $h_{ab}=\gtnoll_{ab}$ for any
 admissible value of $n_0$, coincide with
 $\Tr{\left\{\left[\mathbf{e}-\ts\frac13(\Tr{\mathbf{e}})\mathbbm{1}\right]^2\right\}}$
 up to second order in $\mathbf{e}$.} 
\item{$s^2$ should be positive
 definite in the sense that it should be be strictly positive except
 at zero shear where it should vanish.}  
\item{$s^2$ should be an
 exact invariant of $\pmb{\eta}$.}
\end{itemize}
According to \refeq{seq} a natural choice is to simply replace the
locally defined shear matrix $\mathbf{e}$ in \refeq{s2eq} by either of the
globally defined constant volume shear matrices
$\mathbf{s}$ or $\mathbf{\hat{s}}$, i.e.\ by either
\begin{equation}\lbeq{s2g}
  \Tr{\left\{ \left[ \mathbf{s}-\ts\frac13(\Tr{\mathbf{s}})\mathbbm{1}
  \right]^2 \right\}} = \ts\frac14\Tr{\left\{ \left[ 
  \pmb{\eta}-\ts\frac13(\Tr{\pmb{\eta}})\mathbbm{1} \right]^2 \right\}} =
  \ts\frac14\,n^{-4/3}\Tr{\left\{ \left[
  \mathbf{k}-\ts\frac13(\Tr{\mathbf{k}})\mathbbm{1} \right]^2 \right\}}
\end{equation}
or
\begin{equation}\lbeq{s2ginv}
  \Tr{\left\{ \left[ \mathbf{\hat s}-\ts\frac13(\Tr{\mathbf{\hat
  s}})\mathbbm{1} \right]^2 \right\}} = \tsfrac14\Tr{\left\{ \left[ 
  \pmb{\eta}^{-1}-\ts\frac13(\Tr{\pmb{\eta}^{-1}})\mathbbm{1} \right]^2
  \right\}} = \tsfrac14\,n^{4/3}\Tr{\left\{ \left[ 
  \mathbf{k}^{-1}-\ts\frac13(\Tr{\mathbf{k}^{-1}})\mathbbm{1} \right]^2
  \right\}}
\end{equation}
Indeed, the latter is the choice made by Carter and Quintana. However,
we will choose a different definition. Both of the above expressions
have rather cumbersome fractional power dependences on the linear
particle densities $n_\mu$. We found this to lead to unnecessarily
inconvenient formulae when using one of those in practice. The
fractional powers are clearly due to the fact that the invariants are
of second degree in $\pmb{\eta}$ or $\pmb{\eta}^{-1}$ whereas we are
in a three-dimensional space. This suggests that a third degree
invariant should be more suited from this point of view. The following
choice turns out to be a suitable candidate,
\begin{equation}\lbeq{s2our}
  s^2 = \tsfrac1{36}\left[(\Tr \pmb{\eta})^3 - \Tr (\pmb{\eta}^3)-24\right] 
  = \tsfrac1{36}\,n^{-2}
 \left[(\Tr \mathbf{k})^3 - \Tr (\mathbf{k}^3)-24\det \mathbf{k} \right] 
\end{equation}
Remarkably, this invariant looks precisely the same expressed in
$\pmb{\eta}^{-1}$, \ie it is symmetric under $\pmb{\eta}
\leftrightarrow \pmb{\eta}^{-1}$. It is not at all obvious that it
fulfills the first two of the three requirements listed above. That
the first requirement holds true can be proved in a straightforward
manner. In order to show that our $s^2$ fulfills the second requirement
of positive definiteness we simply write down the invariant in terms
of the linear particle densities. It is then given by
\begin{equation}
\begin{split}
s^2&=\frac1{36\,(n_1\,n_2\,n_3)^2} \left[\,
  (n_1^{\,\,2}+n_2^{\,\,2}+n_3^{\,\,2})^3 -
  (n_1^{\,\,6}+n_2^{\,\,6}+n_3^{\,\,6}) - 24\,(n_1\,n_2\,n_3)^2
  \,\right] \\ 
&= \frac1{12}\left[\, \left(\frac{n_1}{n_2}\right)^{\!\!2} +
  \left(\frac{n_2}{n_1}\right)^{\!\!2} +
  \left(\frac{n_2}{n_3}\right)^{\!\!2} +
  \left(\frac{n_3}{n_2}\right)^{\!\!2} +
  \left(\frac{n_3}{n_1}\right)^{\!\!2} +
  \left(\frac{n_1}{n_3}\right)^{\!\!2} - 6 \,\right]  \\   
&= \frac1{12}\left[\, \left(\frac{n_1}{n_2} - \frac{n_2}{n_1} \right)^{\!\!2} +
  \left( \frac{n_2}{n_3} - \frac{n_3}{n_2} \right)^{\!\!2} + \left(
  \frac{n_3}{n_1} - \frac{n_1}{n_3} \right)^{\!\!2} \,\right].
\end{split}
\end{equation}
Clearly, the last expression shows that our $s^2$ is positive except
at $n_1 = n_2 = n_3$, \ie at vanishing shear $\pmb{\eta} = \mathbbm{1}$. Let us
finally consider the degenerate case of two equal particle
densities. In section \ref{sec:deg} we saw that setting $n_2=n_3$ gives 
\begin{equation}
  s^2 = \tsfrac16(z^{-1} - z)^2
\end{equation}
where $z=n_1/n_2$. Had we instead chosen the definition of $s^2$ to be
given by eq.\ \refeq{s2g}, the corresponding expression would be
\begin{equation}
  s^2 = \tsfrac16 z^{2/3}(z^{-1} - z)^2, 
\end{equation}
or, using eq.\ \refeq{s2ginv},
\begin{equation}
  s^2 = \tsfrac16 z^{-2/3}(z^{-1} - z)^2.
\end{equation}
For this degenerate case it is clear that the choice \refeq{s2our}
has the simplest dependence on the particle densities.

 
\newpage
\pagebreak[4]

\begin{figure}[!hb]
  \centering\begin{minipage}[t]{0.78\linewidth} \centering
 \vspace{0pt} \centering 
 \includegraphics[width=0.7\textwidth,
 angle=270]{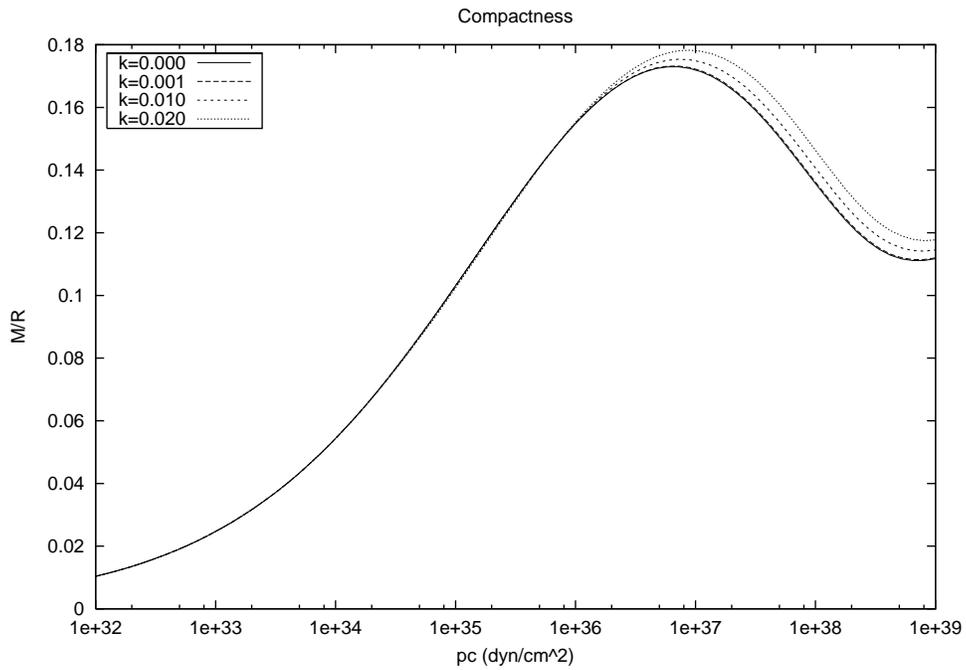}
 \caption{Compactness as a function of central pressure for models
 with a solid core.} 
\label{fig:core2} 
\end{minipage}
\end{figure}

\end{document}